\documentclass[superscriptaddress,pop,english,11pt,reqno]{revtex4-1}
\usepackage[titletoc,toc,title]{appendix}
\usepackage{fullpage}
\usepackage{amsmath}
\usepackage{bm}
\usepackage{amssymb}
\usepackage{amsfonts}
\usepackage{mathtools}
\usepackage{graphicx,color,psfrag}
\usepackage{epsfig}
\usepackage{subfig}

\usepackage[usenames,dvipsnames]{pstricks}
\usepackage{epsfig}
\usepackage{pst-grad} 
\usepackage{pst-plot} 

\usepackage{soul}
\sethlcolor{yellow}

\usepackage{algpseudocode}

\DeclareMathAlphabet{\mathpzc}{OT1}{pzc}{m}{it}

\usepackage{hyperref}
\usepackage{booktabs} 
\usepackage{array}    
\usepackage{paralist} 
\usepackage{verbatim} 
\usepackage{caption}
\usepackage{fancyhdr}
\usepackage{graphicx} 

\newcommand{\zebra}{\ensuremath {\breve{\text z}} }

\begin{document}

\title{Conservative Algorithms for Non-Maxwellian Plasma Kinetics}
\author{Hai P. Le}
\thanks{Corresponding author. Electronic mail: hle@llnl.gov}
\affiliation{Lawrence Livermore National Laboratory, Livermore, CA 94551, USA}
\author{Jean-Luc Cambier}
\affiliation{Air Force Office of Scientific Research, Arlington, VA 22203, USA}

\date{\today}

\begin{abstract}

We present a numerical model and a set of conservative algorithms for Non-Maxwellian plasma kinetics with inelastic collisions. These algorithms self-consistently solve for the time evolution of an isotropic electron energy distribution function interacting with an atomic state distribution function of an arbitrary number of levels through collisional excitation, deexcitation, as well as ionization and recombination. Electron-electron collisions, responsible for thermalization of the electron distribution, are also included in the model. The proposed algorithms guarantee mass/charge and energy conservation in a single step, and is applied to the case of non-uniform gridding of the energy axis in the phase space of the electron distribution function. Numerical test cases are shown to demonstrate the accuracy of the method and its conservation properties.

\end{abstract}

\maketitle

\section{Introduction}
 \label{sec:intro}
The relaxation kinetics of a non-equilibrium and high-energy density plasma are of great importance for a number of applications, such as inertial confinement fusion, laser-produced plasma, material processing, space propulsion and micro-plasma discharges for flow control \cite{betti_inertial-confinement_2016,vinko_creation_2012, lieberman_principles_2005, biberman_kinetics_2012, gundersen_gas-phase_1991}. As the time scales of interest are reduced and the electrons have higher energies, non-equilibrium effects become more severe. High electron energies allow rapid excitation and ionization of the atomic or molecular states, including multiple ionizations. Prediction of the radiative properties, an increasingly important component of the plasma at high temperatures, requires a Collisional-Radiative (CR) description of the plasma, where detailed population of electronic states is calculated and hence allowed to deviate from a Boltzmann distribution\cite{ralchenko_modern_2016}. 
The time-resolved kinetics of the Atomic State Distribution Function (ASDF) for a CR plasma can be complex, even when the Electron Energy Distribution Function (EEDF) is at equilibrium (Maxwellian); a practical solution of the system may require the grouping of quantum levels into various ``states'' obtained by averaging over the fundamental constituent levels\cite{scott_advances_2010,le_complexity_2013}. The difficulty increases considerably when the EEDF itself is not in equilibrium; this can occur due to high accelerating fields, strong coupling with inelastic collisions and short time scales\cite{oxenius_kinetic_1986}. There are two basic methods available to solving the non-equilibrium EEDF kinetics: particle-based methods or discretized phase-space. 
Since we are interested here only in the collisional terms of the kinetic equation, with an emphasis on the inelastic collisions, we will not discuss the respective merits of either for transport. When a particle method such as the Particle-in-Cell (PIC) method is used, the collisions are usually treated with a Monte-Carlo Collision (MCC) algorithm. 
In that scheme, inelastic collision events are selected at random, according to the respective probabilities of occurrence during a given time-step $\Delta t$. This standard approach has been extensively studied\cite{nanbu_simple_1994,yan_analysis_2015}, and it has the advantage that high degrees of non-equilibrium (e.g. anisotropic distributions, beams) can be easily treated. However, statistics for the EEDF can be very poor, and are improved only at a considerable computational expense by increasing the number of pseudo-particles \cite{yan_analysis_2015}. 
Here, we are interested in the second approach, i.e. phase-space discretization, which is especially useful for conditions not too far from equilibrium, in which case useful approximations can be made. For example, electron-electron ($ee$) collisions can be modeled by a Fokker-Planck (FP) collision operator, i.e., a convection-diffusion equation in velocity space \cite{rosenbluth_fokker-planck_1957}; this approach yields good solutions when small angle scattering collisions dominate\cite{epperlein_implicit_1994,yoon_fokker-planck-landau_2014,taitano_mass_2015}. However for inelastic collisions, the change in energy is large and the transfer terms are non-local, i.e. the process transfers a number of electrons from a phase-space cell to other cells which can be far (non-neighboring) from the initial cell -- a ``jump'' process. 
In this case, one must consider the full Boltzmann collision operator, the solution of which leads to a set of master equations for the population of each cell in phase space\cite{rockwood_elastic_1973,morgan_elendif:_1990,marchand_simulation_1991,colonna_boltzmann_2008,dangola_efficient_2010}. This approach can be much faster than MCC and does not suffer from statistical noise; it is, however, challenging for multi-dimensional phase space (6D in the most general case). If the EEDF is nearly isotropic, the number of dimensions can be reduced and the problem becomes more easily solvable \cite{shkarofsky_cartesian_1963}.

In the present work, we are solving the coupled ASDF-EEDF in a non-equilibrium situation, i.e. for a non-relativistic, non-Maxwellian case of CR kinetics. The inelastic collisions between electrons and atomic states are examined, for an arbitrary number of excited levels. Both excitation and ionization are considered, as well as the reverse processes of deexcitation and recombination. Since the energy exchange from these inelastic collisions can be large, it is critically important to guarantee energy conservation. The electron-electron collisions are also included, as they are the main thermalization mechanism in a highly ionized plasma. The importance of energy conservation was emphasized in previous work of many authors\cite{rockwood_elastic_1973,morgan_elendif:_1990,colonna_boltzmann_2008,dangola_efficient_2010}. A brief description and differences among these methods was summarized in D'Angola et al.\cite{dangola_efficient_2010} In the work of Rockwood\cite{rockwood_elastic_1973}, the author formulated an energy conserving discretization of the electron-electron collision operator using a finite volume method on an uniform energy grid. The essence of Rockwood method is to alter the collisional rates to satisfy energy conservation. In his original work, Rockwood used an explicit method to integrate the discretized equations for electron-electron collision term. Although energy conserving, this treatment makes the numerical time steps very restrictive. Morgan and Penetrante\cite{morgan_elendif:_1990} later improved upon Rockwood's work by using a semi-implicit treatment of the electron-electron collision term. Although time step restriction can be relaxed, the semi-implicit aspect of the time integration breaks the energy conservation property of the Rockwood's formulation. A fully implicit treatment was later implemented in the work of Colonna and Capitelli\cite{colonna_boltzmann_2008} and D'Angola et al.\cite{dangola_efficient_2010}, leading to a non-linear system of equations, which are then solved by an iterative method. It is worth mentioning that all of the previous methods described here made the assumption that the time variation of the atomic kinetics is slower compared to the EEDF, such that the inelastic collision operator becomes linear. This is not the case in our work since we consider a fully-coupled system with self-consistent coupling between the EEDF and the ASDF. In addition, the presence of three-body recombination also breaks the linearity of the inelastic collision operator. {A review of self-consistent approaches for CR models can be found in \cite{colonna_self-consistent_2016}}.

In this paper, we consider a fundamentally different approach to formulate an energy conserving discretization of the kinetic equations. Our approach relies on a high-order local moment expansion of the distribution, which effectively gives us an additional degree of freedom to describe energy transfer among energy bins. Since the bin energies are also part of the solution variables, energy conservation can be easily achieved without the need to modify the rates. The advantages of our method are that it can be generalized to other collision operators, the rates are physical, and the extension to the case of a non-uniform energy grid is straightforward. We will show later that the discretization conserves energy exactly for both inelastic and electron-electron collisions. Here only an isotropic EEDF is considered. Generalization to the anisotropic case is possible, although beyond the scope of this work. For example, one can rely on a multi-term expansion, e.g., spherical harmonic or Cartesian tensorial expansion \cite{shkarofsky_cartesian_1963}, of the full distribution function, which includes also the anisotropic component. The inelastic collision operators described in this work could be applied directly to the isotropic component while their impact on other (high-order) terms can be assumed negligible\cite{hagelaar_solving_2005}.

This paper is organized as follows. In section \ref{sec:formulation}, we formulate the kinetic equation for the EEDF and the master equations for the atomic states due to both inelastic and elastic collision processes. We also demonstrate that for a standard discretization procedure, energy conservation is not guaranteed. In section \ref{sec:expand}, the moment expansion approach is introduced and applied to the inelastic collision operators, which allows us to derive numerical algorithms that conserves energy exactly. The discretization of the electron-electron collision term is described in section \ref{sec:thermalization} with a standard discontinuous Galerkin method. We propose to augment the standard discontinuous Galerkin source terms by an equivalent flux form to achieve exact energy conservation. Finally, we present a series of numerical tests in section \ref{sec:numerics} to verify and demonstrate the capabilities of the proposed algorithms, and conclude by providing a summary and prospects for future work in section \ref{sec:conclusion}. We also demonstrate the H-theorem for the case of excitation and ionization collisions in Appendix \ref{app:h-theorem}.

\section{Mathematical formulation}
\label{sec:formulation}
\subsection{Kinetic equations}
We consider the collisional kinetics of an isotropic, quasi-neutral and homogeneous plasma free of external fields. The plasma consists of atoms (neutrals and ions) and free electrons. The atoms are assumed to be static, i.e., their velocity changes due to collisions are negligible and the relative velocity between an atom and an electron is simply the electron velocity; this is a reasonable approximation since the the atoms are much heavier than the free electrons. Furthermore, the time scale of interest in the current study is on the order of electron collision time, which is usually much shorter than time scale of the atoms.

In thermodynamic equilibrium, denoted by a superscript $(\star)$, the EEDF follows an isotropic Maxwellian distribution:
\begin{align}\label{eq:Maxw}
f^\star (\varepsilon) = 2 N_e (\pi T^3)^{-1/2} \varepsilon^{1/2} e^{-\varepsilon/T}
\end{align}
where $N_e$ is the electron number density and $T$ is the equilibrium temperature (for brevity, the Boltzmann constant is omitted throughout the text, so temperature has the same unit as energy). The distribution function $f$ has units of cm$^{-3}$-eV$^{-1}$, and in the general case, the total number density can be determined directly by integrating $f$ over all energy values, i.e., $N_e = \int_0^{\infty} f \, d\varepsilon$. For the ASDF, the Boltzmann relation applies to any two levels $l$ and $u$ within the same ion:
\begin{equation}\label{eq:Boltz}
\left(\frac{N_u}{N_l}\right)^\star = \frac{g_u}{g_l}e^{-\Delta E_{lu}/T}
\end{equation}
where $g_l$ and $g_u$ are the atomic level degeneracies, and $\Delta E_{lu} = E_u - E_l$ is the energy gap. In addition, the Saha equation specifies an equilibrium relation between two levels of two adjacent ions: 
\begin{equation}\label{eq:Saha}
\left(\frac{N^+_u N_e}{N_l}\right)^\star = \frac{2 g^+_u}{g_l} \lambda_e^{-3} e^{-I_l/T}
\end{equation}
where $\lambda_e = \frac{h}{\left( 2 \pi m_e T \right)^{1/2}}$ is the electron thermal de Broglie wavelength, and $I_l$ is the ionization energy from $l$ to $u$. Here we use a superscript $(+)$ to indicate that the state $u$ is ionized once further, i.e. $Z_u\!=\!Z_l\!+\!1$.

In a nonequilibrium system, the time evolution of the EEDF is governed by means of a kinetic equation, and the ASDF by a system of rate equations, one for each level $k$:
\begin{subequations}
\begin{align}
\label{eq:BFP}
\partial_t f (\varepsilon) &= Q^\text{ED}_e + Q^\text{IR}_e - \partial_\varepsilon J_{ee}\\
\label{eq:master_eqs}
\partial_t N_k &= Q^\text{ED}_k + Q^\text{IR}_k
\end{align}
\end{subequations}
The first and second terms on the right hand side of (\ref{eq:BFP}) are the Boltzmann collision operators for the electrons due to excitation/deexcitation (ED) and ionization/recombination (IR). The third term is a Fokker-Planck (FP) term responsible for describing $ee$ collisions. Similarly, we also have two corresponding terms in (\ref{eq:master_eqs}) for the time rate of change of the atomic states due to ED and IR. Since our focus is on a conservative treatment of the collisions, we assume that the plasma is free of the electromagnetic field. This assumption simplifies the electron kinetic equation, but can also be considered by adding an extra term to eq.~(\ref{eq:BFP}) (see eq. (2) of \cite{rockwood_elastic_1973}). Note that we are not considering here the radiative transitions, which will be examined in the future.

\begin{figure}
	\centering
    \subfloat[]{
    	\includegraphics[scale=.8]{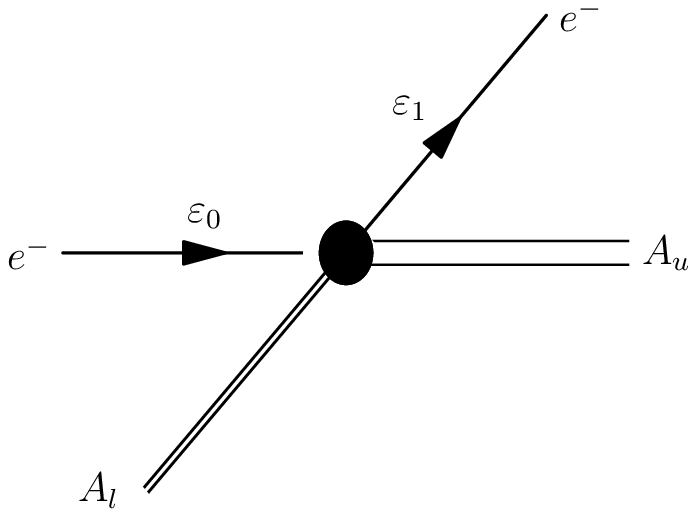}
        
        \label{fig:coll_schematic_a}
    }
    \subfloat[]{
    	\includegraphics[scale=.8]{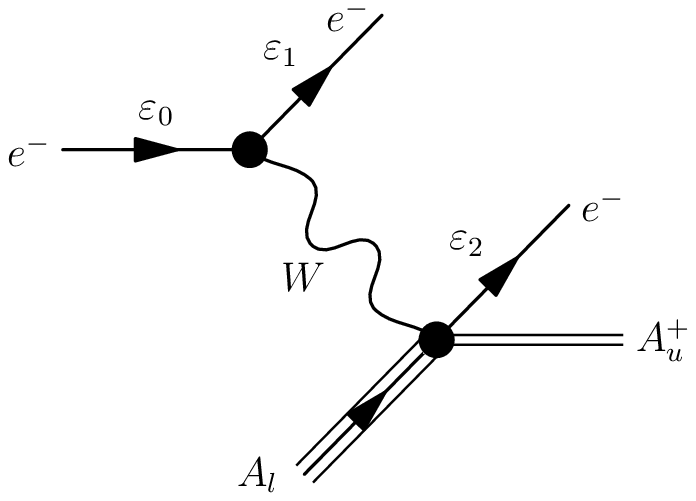}
    	
     	\label{fig:coll_schematic_b}
    }
    \caption{\footnotesize{Schematics of (a) excitation and (b) ionization processes. In the case of ionization, the energy exchanged $W$ can take a range of values, while it is constrained to a single value (the excitation energy) in the first case.}}
    \label{fig:coll_schematic}
\end{figure}

Let us examine these collision operators in more detail starting with the inelastic collisions (ED and IR). For simplicity, we write the time rate of change of the EEDF and the atomic densities due to a single transition (ED or IR) between a unique set of initial and final atomic states, denoted as $\left. \partial_t f(\varepsilon) \right]^{\text{ED}}_{lu}$. The total rate of change can be obtained by summing over all transitions, e.g., $Q^\text{ED}_e = \sum_{l} \sum_{u>l} \left. \partial_t f(\varepsilon) \right]^{\text{ED}}_{lu}$. Consider now a single excitation transition between a free electron and an atom, the result of which leads to an excitation of the atom from a lower state $l$ to an upper state $u$ ($l<u$). We denote this transition as $(\varepsilon_0,l ; \varepsilon_1, u)$ where $\varepsilon_0$ and $\varepsilon_1$ are the initial and final energies of the electron respectively ($\varepsilon_0 > \varepsilon_1$). A schematic of this process is given in Fig. \ref{fig:coll_schematic_a}. The reverse process of this is a deexcitation from $u$ to $l$ of the form $(\varepsilon_1 , u ; \varepsilon_0 , l)$. The time rates of change of the EEDF and ASDF due to this transition are:
\begin{subequations}
\begin{align}
\label{eq:ed_f}
\left. \partial_t f(\varepsilon) \right]^{\text{ED}}_{lu} = & \int_{\Delta E_{lu}}^\infty \left( \delta_1 - \delta_0 \right) \left[ N_l f (\varepsilon_0) v_0 \sigma^{\text{exc}}  (\varepsilon_0 ,l;\varepsilon_1 ,u) - N_u f (\varepsilon_1) v_1 \sigma^{\text{dex}} (\varepsilon_1 , u; \varepsilon_0, l) \right] \, d \varepsilon_0\\
\label{eq:ed_nl}
\left. \partial_t N_l \right]^{\text{ED}}_{lu} = & \int_{\Delta E_{lu}}^\infty\left[ -N_l f (\varepsilon_0) v_0 \sigma^{\text{exc}}  (\varepsilon_0 ,l;\varepsilon_1 ,u)  + N_u f (\varepsilon_1) v_1 \sigma^{\text{dex}} (\varepsilon_1 , u; \varepsilon_0, l) \right] \, d \varepsilon_0\\
\label{eq:ed_nu}
\left. \partial_t N_u \right]^{\text{ED}}_{lu} = &-\left. \partial_t N_l \right]^{\text{ED}}_{lu}
\end{align}
\end{subequations}
where $\delta_p \equiv \delta (\varepsilon \!-\! \varepsilon_p)$ is the Dirac delta function, $v_p \!=\! \sqrt{2 \varepsilon_p / m_e}$ ($p\!=\!0,\!1,\!2$), and $\sigma^{\text{exc}}$ and $\sigma^{\text{dex}}$ are the excitation and deexcitation cross sections, respectively. Strictly speaking, since the electron can change direction after the collision, one must use a differential cross section $\frac{d\sigma}{d \mathit{\Omega}}$ which also includes the probability distribution of the scattering angles. However, since we assume that the EEDF is isotropic, the use of a total cross section $\sigma$, i.e., integrated over all solid angles of the scattered electrons, is sufficient. 
; the full expression can be found in Oxenius \cite{oxenius_kinetic_1986}. Note that the integration over $\varepsilon_0$ has a lower limit of the energy gap between the two atomic levels $\Delta E_{lu}$. From energy conservation, we have $\varepsilon_0 = \varepsilon_1 + \Delta E_{lu}$. The integrand in (\ref{eq:ed_f}) can be interpreted as the loss and gain terms of the electrons in an infinitesimally small energy bin center at $\varepsilon_0$ and $\varepsilon_1$. The rate of change of the atomic state densities can then be obtained by integrating the source terms in the square bracket of (\ref{eq:ed_f}) over all energy bins as shown in (\ref{eq:ed_nl}) and (\ref{eq:ed_nu}). In addition, it is straightforward to see that $ \int_0^\infty \left. \partial_t f(\varepsilon) \right]^{\text{ED}}_{lu} \, d\varepsilon = 0$, which confirms that the total electron density remains constant. At thermal equilibrium, the forward and backward rates must be equal according to the principle of detailed balance. Setting the integrand in (\ref{eq:ed_nl}) to zero and using the equilibrium distributions (\ref{eq:Maxw}) and (\ref{eq:Boltz}), we obtain a relation for the cross-sections of the forward (excitation) and reverse (deexcitation) processes, known as the Klein-Rosseland formula~\cite{oxenius_kinetic_1986}:
\begin{equation}\label{eq:KR}
	g_l\,\varepsilon_0 \,\sigma^\text{exc} (\varepsilon_0, l ; \varepsilon_1, u)
  = g_u\,\varepsilon_1 \,\sigma^\text{dex} (\varepsilon_1, u ; \varepsilon_0, l)
\end{equation}

Let us consider now an ionization transition from level $l$ to level $u$. We denote this transition as $(\varepsilon_0, l ; \varepsilon_1, \varepsilon_2, u)$, where $\varepsilon_0$, $\varepsilon_1$ and $\varepsilon_2$ are the initial, scattered and ejected electrons, respectively. A schematic of this process is given in Fig. \ref{fig:coll_schematic_b}. The reverse process of this is a three-body recombination, denoted as $(\varepsilon_1, \varepsilon_2, u ; \varepsilon_0, l)$. The source terms for both of these processes are written as:
\begin{subequations}
\begin{align}
\label{eq:ir_f}
\left. \partial_t f(\varepsilon) \right]^{\text{IR}}_{lu} = & \int \left( - \delta_0 + \delta_1 + \delta_2 \right) \left[ N_l f (\varepsilon_0) v_0 \sigma^{\text{ion}} (\varepsilon_0, l; \varepsilon_1, \varepsilon_2, u) \right. \nonumber \\
& \left. - N_u^+ f (\varepsilon_1) f (\varepsilon_2) v_1 v_2 \sigma^{\text{rec}} (\varepsilon_1, \varepsilon_2, u; \varepsilon_0, l) \right] \delta (\{ \varepsilon \})\, d\varepsilon_0 \, d\varepsilon_1 \, d \varepsilon_2\\
\label{eq:ir_nl}
\left. \partial_t N_l \right]^{\text{IR}}_{lu} = & \int \left[- N_l f (\varepsilon_0) v_0 \sigma^{\text{ion}} (\varepsilon_0, l; \varepsilon_1, \varepsilon_2, u) \right. \nonumber  \\
& \left. + N_u^+ f (\varepsilon_1) f (\varepsilon_2) v_1 v_2 \sigma^{\text{rec}} (\varepsilon_1, \varepsilon_2, u; \varepsilon_0, l) \right] \delta (\{ \varepsilon \})\, d\varepsilon_0 \, d\varepsilon_1 \, d \varepsilon_2 \\
\label{eq:ir_nu}
\left. \partial_t N_u^+ \right]^{\text{IR}}_{lu} =& \left. -\partial_t N_l \right]^{\text{IR}}_{lu}
\end{align}
\end{subequations}
where $\sigma^{\text{ion}}$ and $\sigma^{\text{rec}}$ are the differential ionization and recombination cross sections respectively, and $\delta (\{ \varepsilon \}) = \delta (\varepsilon_0 - \varepsilon_1 - \varepsilon_2 - I_l)$ from energy conservation. Note that the integration over $\varepsilon_0$ starts at the lower threshold of the ionization energy $I_l$. It is common to define $W \!=\! \varepsilon_0 \!-\! \varepsilon_1$ as the energy transfer, such that $W \!=\! \varepsilon_2 \!+\! I_l$. Similarly to the excitation/deexcitation case, it can be shown from the principle of detailed balance that the differential cross-sections for ionization and recombination are related by the so-called Fowler formula~\cite{oxenius_kinetic_1986}.
\begin{equation}\label{eq:Fowler}
g_l \varepsilon_0 \sigma^\text{ion} (\varepsilon_0, l; \varepsilon_1, \varepsilon_2, u) = \frac{16\pi m_e}{h^3}\ g_u                                          \varepsilon_1 \varepsilon_2  \sigma^\text{rec} (\varepsilon_1, \varepsilon_2, u; \varepsilon_0, l)
\end{equation}
The rate equations for the atomic states $(l,u)$ are determined by integrating the source terms in eq.~(\ref{eq:ir_f}) over all initial and final electron energies, as seen in eqs. (\ref{eq:ir_nl}) and (\ref{eq:ir_nu}). Note that the total electron number density can change due to creation and removal of electrons. The total rate of change of the electron density is the same as that of the ion density, i.e., $ \int_0^\infty \left. \partial_t  f \right]^{\text{IR}}_{lu} \, d\varepsilon = \left. \partial_t N_u^+ \right]^{\text{IR}}_{lu}$, following mass and charge conservation.

A proof of the H-theorem for excitation/deexcitation and ionization/recombination is given in appendix \ref{app:h-theorem}, which shows that the entropy production is nonnegative, i.e., $\partial S/ \partial t \ge 0$. Equivalently, one can define a Boltzmann's $H$ function ($\propto -S$) and show that $\partial H/ \partial t \le 0$. For a more in depth discussion, the reader is referred to the work of Yan et al. \cite{yan_analysis_2015} We will demonstrate later in section \ref{sec:acc_tests} that the numerical algorithms proposed in this work are consistent with H-theorem.

The third term on the right hand side of (\ref{eq:BFP}) is the electron-electron collision term, which leads to thermalization of the EEDF. For an isotropic system, this term appears as a divergence of a flux in energy space. In this study, we use the same formulation as Rockwood \cite{rockwood_elastic_1973}, which was derived from the Rosenbluth form of the FP equation\cite{rosenbluth_fokker-planck_1957}. The flux is written as follows:
\begin{subequations}
\label{eq:fluxee}
\begin{align}
J_{ee} &= \gamma \left[ K ( f/2\varepsilon - \partial_\varepsilon f ) - L f \right] \\
\gamma &= \frac{2}{3} \pi e^4 (2/m_e)^{1/2} \ln \varLambda\\
K &= 2 \varepsilon^{-1/2} \int_0^\varepsilon \varepsilon' \, f (\varepsilon') \, d\varepsilon' + 2 \varepsilon \int_{\varepsilon}^{\infty} \varepsilon'^{-1/2} \, f (\varepsilon') \, d\varepsilon'\\
L &= 3 \varepsilon^{-1/2} \int_0^\varepsilon f (\varepsilon') \, d\varepsilon'
\end{align}
\end{subequations}
where $\ln \varLambda$ is the Coulomb logarithm. It must be pointed out that this collision operator conserves the total number density and energy of the electrons; hence the numerical discretization should respect these properties. In addition, one can show that the H-theorem holds for the $ee$ collision operator, that is, the time rate of change of the functional $H \equiv \int_0^\infty f \ln f \, d\varepsilon$ due to this process, is a non-increasing function, i.e., $\left. \partial_t H \right]_{ee} \leq 0$, and the equality holds iff $f$ is a Maxwellian distribution.

\subsection{Standard discretization}
\begin{figure}
\centering
	\subfloat[]{
		\includegraphics[scale=0.7]{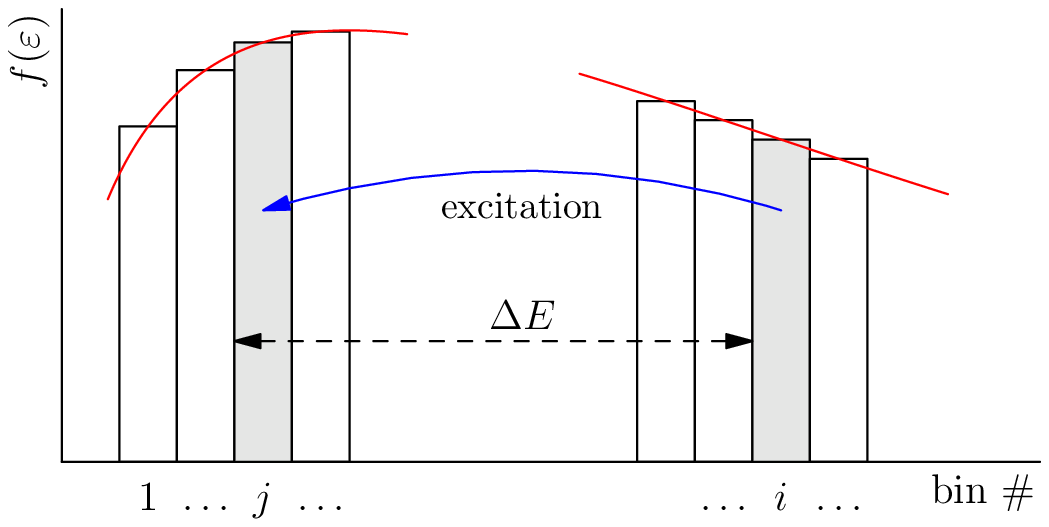}	\label{fig:jump_xd_a}
	}
	\subfloat[]{
		\includegraphics[scale=0.7]{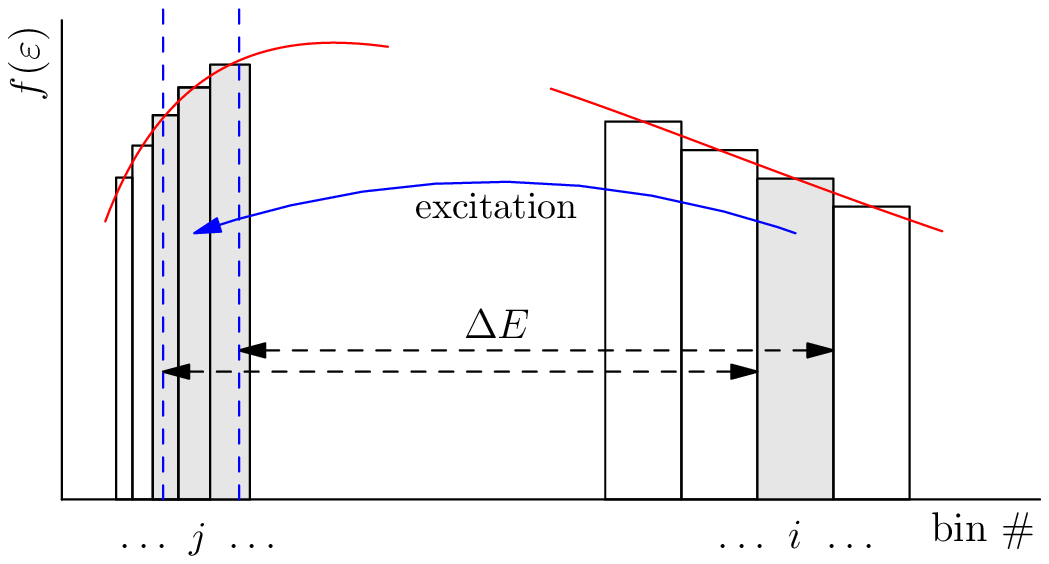}
		\label{fig:jump_xd_b}
	}
		\caption{\footnotesize{Mapping of excitation process from initial bin $[i]$; (a) to a single final bin $[j]$ in the case of exact alignment of energy gap $\Delta E$ with bin boundaries; (b) to a collection of bins $\{[j]\}$, when the threshold $\Delta E$ is not aligned between between bin boundaries, and when $\Delta E < \varepsilon_{i-1/2}$, the lower boundary of bin $[i]$. The dashed vertical blue lines indicate the limits of the initial energy bin $[i]$ translated by the energy gap $\Delta E$.}}
		\label{fig:jump_xd}
\end{figure}

In this section, we briefly describe the standard discretization procedure for inelastic collisions\cite{rockwood_elastic_1973} and show that energy conservation is not guaranteed. The elastic $ee$ collision term is treated separately in section \ref{sec:thermalization}. For simplicity, we only consider excitation/deexcitation collisions as an example; the same argument can be applied to ionization/recombination. Let us discretize the EEDF into $N_b$ bins of constant width $\Delta \varepsilon$. The number density of electrons in bin $i$ is defined as:
\begin{equation}\label{eq:2.15}
\overline{n}_i = \int_{\varepsilon_{i-1/2}}^{\varepsilon_{i+1/2}} f (\varepsilon) \, d\varepsilon 
\end{equation}
where the subscript $i \pm 1/2$ denotes the left and right boundaries of the bin. Consider the case of a single atomic transition $(\varepsilon_0,l ; \varepsilon_1,u)$, and assume that the energy gap is an exact integer multiple of the constant bin width, $\Delta E_{lu} = K \Delta \varepsilon$; in that case, the excitation process is a simple and exact transfer between the initial bin $i$ and the final $j=i-K$ (see Fig. \ref{fig:jump_xd_a}) given that $\varepsilon_{i-1/2} > \Delta E_{lu}$. We can then define an excitation rate coefficient as follows:
\begin{align}
\label{eq:kn_exc}
\overline{k}^{\text{exc}}_{ij} = \frac{ \int_{\varepsilon_{i-1/2}}^{\varepsilon_{i+1/2}} f(\varepsilon_0) v_0 \sigma^{\text{exc}} \, d\varepsilon_0}{\overline{n}_i}
\end{align}
where the subscripts denoting atomic level indices are dropped for brevity. For the reverse process $(\varepsilon_1,u ; \varepsilon_0,l)$, a similar deexcitation rate can be written as:
\begin{align}
\overline{k}^{\text{dex}}_{ji} = \frac{ \int_{\varepsilon_{j-1/2}}^{\varepsilon_{j+1/2}} f(\varepsilon_1) v_1 \sigma^{\text{dex}} \, d\varepsilon_1}{\overline{n}_j}
\end{align}
The time rate of change of the bin densities can be expressed as:
\begin{subequations}
\begin{align}
\frac{d \overline{n}_i}{dt} &= -N_l \overline{n}_i \overline{k}^{\text{exc}}_{ij}  + N_u \overline{n}_j \overline{k}^{\text{dex}}_{ji}\\
\frac{d \overline{n}_j}{dt} &=  N_l \overline{n}_i \overline{k}^{\text{exc}}_{ij}  - N_u \overline{n}_j \overline{k}^{\text{dex}}_{ji}
\end{align}
\end{subequations}
Note that in this case since the bins are perfectly aligned, so we only have one initial and one final bin. With multiple transitions, it is impossible to have perfect alignment of the energy gaps and bin boundaries; each transition will therefore include bin overlaps and, in the case of non-uniform binning, multiple final bins (see Fig. \ref{fig:jump_xd_b}). The strategy to deal with this situation, and the approach in computing the effective rates will be discussed later in section \ref{sec:exc}. Nevertheless, using these effective rates for all the bins, one can construct a system of the form:
\begin{align}
\label{eq:standard_ex}
\frac{d \overline{n}_i}{dt} = a^{\text{exc}}_{ij} \overline{n}_j + b^{\text{dex}}_{ij} \overline{n}_j; \quad i,j=1,\hdots,N_b
\end{align}
where summation over repeated indices is implied. Thus, the terms on the right hand side of eq.~(\ref{eq:standard_ex}) are matrix-vector products, and $a^{\text{exc}}_{ij}, b^{\text{exc}}_{ij}$ are elements of the rate matrix connecting $i$ and $j$, and must contain a sum over all ED transitions. Adding the equations for the ASDF evolution, we obtain an extended set of ODEs describing the collisional kinetics with an overall vector of variables which includes all atomic states ($N_k$) and the set of bin densities ($\overline{n}_i$).

The standard discretization, defined in (\ref{eq:standard_ex}), is not necessarily energy-conserving since it only describes the evolution of the bin densities. Assuming that the average energy of a bin is the same as the value at the center, it is straightforward to see that the total rate of change of the energy due to a transition involving bins $i$ and $j$ is:
\begin{align}
\label{eq:2.44}
\frac{d \overline{e}_i}{dt} + \frac{d \overline{e}_j}{dt} = \frac{d\overline{n}_i}{d t} (\varepsilon_i - \varepsilon_j)
\end{align}
where $\overline{e}_i$ denotes the total energy of the bin and $\varepsilon_i = \frac{\varepsilon_{i-1/2} + \varepsilon_{i+1/2}}{2}$ is the bin center. Eq. (\ref{eq:2.44}) indicates that energy conservation is only satisfied if $\varepsilon_i - \varepsilon_j = \Delta E_{lu}$, i.e., the distance between the bin centers exactly matches the energy gap of the transition. As mentioned before, it is impossible to have bin allignment for multiple transitions, so energy conservation is not guaranteed in general. For example, if we consider a transition between two highly excited states, where the energy gap is much smaller than the grid size ($\Delta E_{lu} \ll \Delta \varepsilon$), it is possible that the initial and final bins are the same ($i=j$). In this case, the standard discretization results in zero changes in both density and energy of the bin, which is obviously incorrect. We emphasize that the non-conservation is primarily due to the lack of sufficient degrees of freedom per bin; therefore, the ability to go beyond a single degree of freedom per bin would allow us to derive an energy-conserving scheme. This is the foundation of the moment expansion method introduced in the next section, which naturally guarantees energy conservation in a single time step regardless of the step size.

\section{Inelastic collisions}
\label{sec:expand}
\subsection{Moment expansion}
\label{sec:expand-1}
In the previous section, we have demonstrated a straightforward method of constructing master equations for the bins of the EEDF. These determine for the number densities of electrons within each bin, and are therefore ``mass''-conserving. We now describe a method to extend the conservation law to include energy, via a local high-order expansion.
Consider now the discretization of the EEDF into $N_b$ bins of \textit{variable} width $\Delta_i\varepsilon$ ($i=1,\hdots,N_b$), we use an expansion of the EEDF in \textit{each bin} $i$ in terms of an orthonormal set of basis functions:
\begin{equation}\label{eq:expansion}
	f_i(\varepsilon)=\sum_{p=0}^{N_p}a^i_p U_p(\varepsilon); \quad i=1,\hdots,N_b
\end{equation}
such that $\int_{-1}^{+1}U_p(z)\, U_q(z)\, dz = \delta_{pq}$ with a change of variable $\varepsilon= \varepsilon_i +\frac{\Delta_i\varepsilon}{2} z$. The orthogonality yields
\begin{equation}\label{eq:3.5}
	a_p^i = \int_{-1}^{+1} dz f_i(z) U_p(z)
\end{equation}
Hereafter to simply the expressions, we shall occasionally interchange the variables $z$ and $\varepsilon$, with the understanding that the latter implies the energy within a specified bin. We will choose the normalized Legendre polynomials for the basis functions, i.e., $U_p(z) = \sqrt{p\!+\!\frac{1}{2}}\cdot P_p(z)$ where $P_p(z)$ is the regular Legendre polynomial. 
It is straightforward to show that the density of a bin, defined in (\ref{eq:2.15}), is:
\begin{align}\label{eq:bin_density}
\overline{n}_i = \int_{\varepsilon_{i-1/2}}^{\varepsilon_{i+1/2}} f(\varepsilon) \, d\varepsilon = \frac{\Delta_i\varepsilon}{\sqrt{2}}\cdot a_0^i
\end{align}
where the only contribution is due to the lowest-order polynomial coefficient. Similarly, the energy of a bin is computed from the first moment of $f$ as follows:
\begin{equation}\label{eq:3.11}
	\overline{e}_i= \int_{\varepsilon_{i-1/2}}^{\varepsilon_{i+1/2}} \varepsilon \, f(\varepsilon) \, d\varepsilon =\overline{n}_i \varepsilon_i + \frac{(\Delta_i\varepsilon)^2}{2\sqrt{6}} a_1^i
\end{equation}
Computing higher-order moments leads to a recursive series where each moment of order $p$ can be written as a combination of the moments of lower order and the coefficient $a_p$. Eq. (\ref{eq:3.11}) indicates that energy conservation can be satisfied if the expansion in (\ref{eq:expansion}) is carried to at least first order. Although a higher-order expansion may lead to more accurate representations of the EEDF, this is not necessarily advantageous since the computational requirements increase with the maximum order of the basis functions. In addition, the higher-order moments lose physical meaning; for example, the second-order moment could be associated with the ``temperature'' within the bin (i.e. the energy fluctuations about the mean), but there is no conservation law associated with these higher-order quantities. 
{
Although we presently restrict our numerical implementation to first order, the discretization described in the next section is generalized to expansions of arbitrary order.}

\subsection{Excitation and deexcitation}
\label{sec:exc}
We now describe the procedure to construct elementary rates for the moment variables described above starting with excitation and deexcitation. Assuming the initial electron to have an energy $\varepsilon_0 \in [\varepsilon_{i\!-\!1/2},\varepsilon_{i\!+\!1/2}]$, it is left with an energy $\varepsilon_1\!=\!\varepsilon_0 \!-\!\Delta E$ after an excitation collision, where $\Delta E$ is the energy gap of the transition. Since we focus on a single transition, the atomic level indices are omitted. Two complications immediately arise: first, the threshold energy $\Delta E$ is not necessarily aligned with a bin boundary. As mentioned earlier, even if that was so for a single transition, it is highly impractical to attempt bin alignment for a large number of inelastic transitions. 
Second, the bin widths are not necessarily constant; given the variations of the cross-sections and EEDF, it may in fact be preferable to have variable bin widths. Therefore, we consider the most general case of a mapping from an initial bin $i$ to a \emph{set} of final bins $\{[j]; j=m,...,M\}$, as shown in Fig. \ref{fig:jump_xd_b}. The threshold energy for excitation may lie anywhere inside that initial bin; thus only a fraction $w_i$  of the bin $i$ may contribute to the rate:
\begin{equation}\label{eq:bin_frac_i} 
w_i = \frac{\varepsilon_{i1}-\varepsilon_{i0}}{\Delta_i\varepsilon}
\end{equation}
where $\varepsilon_{i0},\varepsilon_{i1}$ are respectively the lower and upper limits of the electron's energy in bin $i$, and $\Delta_i\varepsilon\!=\!\varepsilon_{i\!+\!1/2}\!-\!\varepsilon_{i\!-\!1/2}$ is the fixed bin width. Since the excitation proceeds from any energy above the threshold $\Delta E$, the upper limit will always be the upper bin boundary, i.e. $\varepsilon_{i1}\!\equiv\!\varepsilon_{i+1/2}$; the lower limit will be $\varepsilon_{i-1/2}$ \emph{except} when the threshold lies within that bin, in which case $\varepsilon_{i0}\equiv\Delta E$. This bin fraction $w_i$ gets mapped onto an arbitrary number of bins at a lower energy range, as a result of the shift by a constant energy $\Delta E$. Similarly to (\ref{eq:bin_frac_i}), one can define a final bin fraction for each member of the final set $\{[j]\}$:
\begin{equation}\label{eq:bin_frac_j}  
w_j = \frac{\varepsilon_{j1}-\varepsilon_{j0}}{\varepsilon_{i1}-\varepsilon_{i0}}
\end{equation}
Note the denominator in (\ref{eq:bin_frac_j}): the normalization is with respect to \textit{all} mappings - see Fig. \ref{fig:jump_xd_b} - such that $\sum_j(\varepsilon_{j1}-\varepsilon_{j0})=\varepsilon_{i1}-\varepsilon_{i0}$.

Consider now the rate of change in the initial bin variables due to a \textit{single} element of the mapping $\mu: [\varepsilon_{i0},\varepsilon_{i1}]\leftrightarrow [\varepsilon_{j0},\varepsilon_{j1}]$ for an excitation. Integrating eq. (\ref{eq:ed_f}) over cell $i$ we obtain:
\begin{align}\label{eq:exc_zi} 
\frac{\Delta_i \varepsilon}{2} \left. \frac{d a_p^i}{dt} \right]^{\text{exc}}_{\mu} &= - N_l \int_{\varepsilon_{j0}+ \Delta E}^{\varepsilon_{j1} + \Delta E}  U_p (\varepsilon_0) f_i (\varepsilon_0) k^{\text{exc}} (\varepsilon_0) \, d \varepsilon_0
\end{align}
where $ k^{\text{exc}} (\varepsilon_0) \! = v_0 \sigma^{\text{exc}} (\varepsilon_0)$. Since we only consider a single element of the mapping, the integration on the right hand side is only over the portion of $i$ which gets mapped to $j$. For convenience, let us define a new set of variables $\zebra_p$, which are renormalized values of the expansion coefficients $a_p^i$:
\begin{align}
\zebra_p (i) = \frac{\Delta_i \varepsilon}{2} a_p^i
\end{align}
These are also related to the local (bin-wise) moments of the distribution. In particular, we have:
\begin{subequations}
\begin{align}\label{eq:3.17}
\overline{n}_i &= \sqrt{2}\cdot\zebra_0(i)\\
\overline{e}_i &= \overline{n}_i \varepsilon_i+\sqrt{\frac{2}{3}}\frac{\Delta_i\varepsilon}{2}\zebra_1(i)
\end{align}
\end{subequations}
The rate of change of this new set of variables can be rewritten in the following form by introducing the expansion (\ref{eq:expansion}) into the right hand side of (\ref{eq:exc_zi}) and using a change of variable.
\begin{align}
\label{eq:exc_zi2}
\left. \frac{d \zebra_p (i)}{dt} \right]^{\text{exc}}_{\mu} &= - \frac{2}{\Delta_i \varepsilon} N_l \sum_q \zebra_q (i) \int_{\varepsilon_{j0}}^{\varepsilon_{j1}}  U_p (\varepsilon_0) U_q (\varepsilon_0) k^{\text{exc}} (\varepsilon_0) \, d \varepsilon_1
\end{align}
The integral over the bin-width can be estimated using a $N_g$-point Gaussian quadrature
\begin{equation}\label{eq:3.14}
\int_a^b h (\varepsilon) \, d\varepsilon \equiv \frac{b-a}{2}\sum_{n=1}^{N_g} \omega_n h(\varepsilon_n)
\end{equation}
with $\omega_n$ the weights, to yield:
\begin{equation}\label{eq:3.15}
	\int_{\varepsilon_{j0}}^{\varepsilon_{j1}} U_p(\varepsilon_0) U_q(\varepsilon_0) k_{ij}^\text{exc}(\varepsilon_0) \, d\varepsilon_1 \simeq \frac{\varepsilon_{j1} - \varepsilon_{j0}}{2} \sum_{n[j]} \omega_n U_p(\varepsilon_{i,n}) U_q(\varepsilon_{i,n}) k^\text{exc} (\varepsilon_{i,n})
\end{equation}
where $n[j]$ denotes the quadrature point $n$ with energy $\varepsilon_{j,n}$ lying within the interval $[\varepsilon_{j0},\varepsilon_{j1}]$ and $\varepsilon_{i,n} \!=\! \varepsilon_{j,n} \!+\! \Delta E$. Here $U_p(\varepsilon_{i,n})$ should be understood as the $p$-th order polynomial function of cell $i$ evaluated at $\varepsilon_{i,n}$.
Inserting (\ref{eq:3.15}) into (\ref{eq:exc_zi2}) and using $\varepsilon_{j1} \!-\! \varepsilon_{j0} \!=\! w_j w_i \Delta_i \varepsilon$, we obtain:
\begin{align}
\left. \frac{d \zebra_p (i)}{dt} \right]^{\text{exc}}_{\mu} &= - N_l \sum_q \zebra_q (i) w_i w_j \sum_{n[j]} \omega_n  U_p (\varepsilon_{i,n}) U_q (\varepsilon_{i,n}) k^{\text{exc}} (\varepsilon_{i,n})
\end{align}
We can now define an elementary rate for $i$ due to each mapping $\mu$ as:
\begin{align}
\label{eq:rate_mui}
\alpha_{\mu,i}^{pq} = w_i w_j \sum_{n[j]} \omega_n  U_p (\varepsilon_{i,n}) U_q (\varepsilon_{i,n}) k^{\text{exc}} (\varepsilon_{i,n})
\end{align}
such that the rate of change of the coefficient $\zebra_p (i)$ can be written in the following form:
\begin{align}
\left. \frac{d \zebra_p (i)}{dt} \right]^{\text{exc}}_{\mu} = -N_l \sum_q \zebra_q (i) \alpha_{\mu,i}^{pq}
\end{align}
The same procedure can be done for the rate of change of the coefficients of bin $j$, leading to:
\begin{align}
\left. \frac{d \zebra_p (j)}{dt} \right]^{\text{exc}}_{\mu} = N_l \sum_q \zebra_q (i) \alpha_{\mu,j}^{pq}
\end{align}
where the elementary rate for $j$ is defined as:
\begin{align}
\label{eq:rate_muj}
\alpha_{\mu,j}^{pq} = w_i w_j \sum_{n[j]} \omega_n  U_p (\varepsilon_{j,n}) U_q (\varepsilon_{i,n}) k^{\text{exc}} (\varepsilon_{i,n})
\end{align}
This result is very similar to eq.~(\ref{eq:rate_mui}), except that the basis function $U_p$ is evaluated at $\varepsilon_{j,n}$, i.e., the quadrature point $n$  in $[\varepsilon_{j0}, \varepsilon_{j1}]$. It is interesting to see that $\alpha_{\mu,i}^{0q} = \alpha_{\mu,j}^{0q}$, which is a statement of mass conservation. Energy conservation can also be shown, although less trivial, such that:
\begin{align}
\frac{d \overline{e}_i}{dt} + \frac{d \overline{e}_j}{dt} = \Delta E \frac{d \overline{n}_i}{dt}
\end{align}

For the reverse process of deexcitation, one can follow the same procedure and construct a different set of mapping from low energy bins to high energy bins. The rate equations for the bin variables can be derived similarly. There is, however, another option, which is to use the same mapping for both forward and backward processes. We have implemented both cases and observed little difference in the results. However, using the same mapping reduces the overall amount of memory and work required, and provides a more exact approach when introducing detailed balance relationships. Therefore, we describe below only one approach, the case of identical mapping for forward and reverse processes. The difference is that now $j$ refers to the initial and $i$ to the final bin. Let us write now the rate of change of the bin variables due to deexcitation for the same mapping $\mu: [\varepsilon_{i0},\varepsilon_{i1}]\leftrightarrow [\varepsilon_{j0},\varepsilon_{j1}]$:
\begin{subequations}
\begin{align}
\left. \frac{d \zebra_p (j)}{dt} \right]^{\text{dex}}_{\mu} &= - \frac{2}{\Delta_j \varepsilon} N_u \sum_q \zebra_q (j) \int_{\varepsilon_{j0}}^{\varepsilon_{j1}}  U_p (\varepsilon_1) U_q (\varepsilon_1) k^{\text{dex}} (\varepsilon_1) \, d \varepsilon_1 \\
\left. \frac{d \zebra_p (i)}{dt} \right]^{\text{dex}}_{\mu} &= \frac{2}{\Delta_j \varepsilon} N_u \sum_q \zebra_q (j) \int_{\varepsilon_{j0}}^{\varepsilon_{j1}}  U_p (\varepsilon_0) U_q (\varepsilon_1) k^{\text{dex}} (\varepsilon_1) \, d \varepsilon_1
\end{align}
\end{subequations}
where $k^{\text{dex}} (\varepsilon_1) \!=\! v_1 \sigma^{\text{dex}} (\varepsilon_1)$, and $\sigma^{\text{dex}}$ is related to $\sigma^{\text{exc}}$ via the Klein-Rosseland relation (\ref{eq:KR}). The quadrature is performed again on the low-energy bin $[j]$, which allows us to obtain:
\begin{subequations}
\begin{align}
\left. \frac{d \zebra_p (j)}{dt} \right]^{\text{dex}}_{\mu} &= - \frac{\varepsilon_{j1} - \varepsilon_{j0}}{\Delta_j \varepsilon} N_u \sum_q \zebra_q (j) \sum_{n[j]} \omega_n  U_p (\varepsilon_{j,n}) U_q (\varepsilon_{j,n}) k^{\text{dex}} (\varepsilon_{j,n})\\
\left. \frac{d \zebra_p (i)}{dt} \right]^{\text{dex}}_{\mu} &= \frac{\varepsilon_{j1} - \varepsilon_{j0}}{\Delta_j \varepsilon} N_u \sum_q \zebra_q (j) \sum_{n[j]} \omega_n U_p (\varepsilon_{i,n}) U_q (\varepsilon_{j,n}) k^{\text{dex}} (\varepsilon_{j,n})
\end{align}
\end{subequations}
The ratio of energies can be written as follows:
\begin{equation}\label{eq:4.27} 
\frac{\varepsilon_{j1}-\varepsilon_{j0}}{\Delta_j \varepsilon}=\frac{\Delta_i\varepsilon}{\Delta_j\varepsilon }\cdot
\frac{\varepsilon_{j1}-\varepsilon_{j0}}{\varepsilon_{i1}-\varepsilon_{i0}}\cdot\frac{\varepsilon_{i1}-\varepsilon_{i0}}{\Delta_i\varepsilon} = \frac{\Delta_i\varepsilon}{\Delta_j\varepsilon}w_jw_i
\end{equation}
Hence the elementary rates can be defined as:
\begin{subequations}
\begin{align}
\beta_{\mu,j}^{pq} &= \frac{\Delta_i\varepsilon}{\Delta_j\varepsilon}w_jw_i \sum_{n[j]} \omega_n U_p (\varepsilon_{j,n}) U_q (\varepsilon_{j,n}) k^{\text{dex}} (\varepsilon_{j,n}) \\
\beta_{\mu,i}^{pq} &= \frac{\Delta_i\varepsilon}{\Delta_j\varepsilon}w_jw_i \sum_{n[j]} \omega_n U_p (\varepsilon_{i,n}) U_q (\varepsilon_{j,n}) k^{\text{dex}} (\varepsilon_{j,n})
\end{align}
\end{subequations}
such that the rate of change for $\zebra_p(i)$ and $\zebra_p(j)$ becomes:
\begin{subequations}
\begin{align}
\left. \frac{d \zebra_p (j)}{dt} \right]^{\text{dex}}_{\mu} &= -N_u \sum_q \zebra_q (j) \beta_{\mu,j}^{pq}\\
\left. \frac{d \zebra_p (i)}{dt} \right]^{\text{dex}}_{\mu} &=  N_u \sum_q \zebra_q (j) \beta_{\mu,i}^{pq}
\end{align}
\end{subequations}
In the numerical calculation, the elementary rates $\alpha^{pq}_{\mu}$ and $\beta^{pq}_{\mu}$ are precomputed for all mappings and all ED processes, since they do not change in time for a static energy grid. This is computationally efficient, especially when the number of transitions gets very large.

\subsection{Ionization and recombination}
\label{sec:ion-rec}
\begin{figure}
\centering
		\includegraphics[scale=0.7]{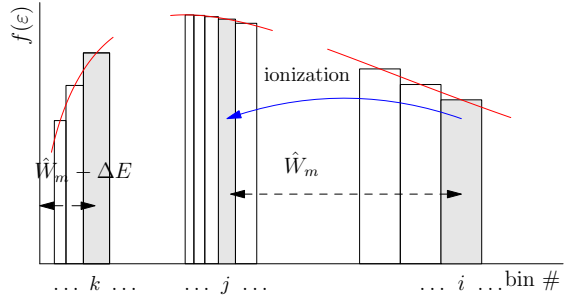}
		\caption{\footnotesize{Mapping of ionization process from initial bin $[i]$ to a pair of final bins $[j,k]$ for a given energy transfer range $[W_m, W_{m+1}]$ using the mid-point average ($\hat{W}_m$) approximation {as will be discussed later in section \ref{sec:ion-rec}}. In this figure, the initial electron is in bin $[i]$, the scattered electron in bin $[j]$, and the ejected electron in bin $[k]$.}}
		\label{fig:jump_ir}
\end{figure}

Let us now look at ionization and three-body recombination, which is more complicated than excitation and deexcitation because there are three electron energy bins involved in the process. In this case, the transferred energy is a variable ranging from the ionization potential to the total energy of the incident electron. Let us denote $i$ to be the initial and $j,k$ the final bin indices. Following the notation in Fig. \ref{fig:coll_schematic_b}, the energies of the incident ($\varepsilon_0 \in [i]$), scattered ($\varepsilon_1 \in [j]$) and ejected ($\varepsilon_2 \in [k]$) electrons are related to the transferred energy $W\!=\!\varepsilon_0\!-\!\varepsilon_1\!=\!\varepsilon_2\!+\!\Delta E$ where $\Delta E$ is now the ionization potential. 
It is then sufficient to consider the variation of the cross section of ionization with respect to $W$, i.e., a singly differential cross section denoted as $\frac{d\sigma^{\text{ion}}}{dW} (\varepsilon_0 ; W)$; the total ionization cross section can then be written as $
\overline{\sigma}^{\text{ion}} (\varepsilon_0) =  \int_{I_l}^{\varepsilon_0} \frac{d \overline{\sigma}^{\text{ion}}}{dW} \, dW$.
In this process, the initial bin fraction $(\varepsilon_{i0},\varepsilon_{i1})$ is mapped onto a set of lower energy bins. The procedure in determining the final bins $(j,k)$ will be described shortly. We first assume that the transferred energy can take the following discrete values:
\begin{equation}\label{eq:4.41}  
		W=\{W_0,\ldots W_m,\ldots W_M\}\qquad \mbox{such that}\qquad W_m=\Delta E\!\cdot\!e^{m\delta W}
\end{equation}
Since $W_0 = \Delta E$ and $W_M=\varepsilon_0$ (the lower and upper limit of the allowed range), the spacing this sequence can be determined as $\delta W = \frac{1}{M}\ln(\frac{\varepsilon_0}{\Delta E})$. 
Note that this geometric sequence has points clustered near the threshold, where the variation of the cross-section is the highest; this is a desirable feature. The corresponding intervals for the final states are:
\begin{subequations}\label{eq:4.43}  
\begin{align}
\varepsilon_1=\varepsilon_0-W &= \{\varepsilon_0\!-\!\Delta E,\ldots,\varepsilon_0\!-\!\Delta E e^{m\delta W},\ldots,0\}\\
\varepsilon_2=W-\Delta E      &= \{0,\ldots,\Delta E(e^{m\delta W}\!-\!1),\ldots,\varepsilon_0\!-\!\Delta E\}
\end{align}
\end{subequations}
Consider now the rate of change of the EEDF for the incident electron:
\begin{equation}\label{eq:4.44}  
	\frac{d}{dt}f_i(\varepsilon_0)=-N_l v_0 f_i(\varepsilon_0) \int_{\Delta E}^{\varepsilon_0}\left(\frac{d\overline{\sigma}^\text{ion}}{dW}\right) \!dW
\end{equation}
Integrating over the initial bin $i$, we obtain the elementary variation of the bin variables - see also eq. (\ref{eq:exc_zi}):
\begin{equation}
\label{eq:ion_zi}
		\left. \frac{d \zebra_p (i) }{dt} \right]^{\text{ion}}_\mu = -N_l \int_{\varepsilon_{i0}}^{\varepsilon_{i1}} d\varepsilon_0 \, U_p(\varepsilon_0) \, v_0 f_i(\varepsilon_0)\int_{\Delta E}^{\varepsilon_0}\left(\frac{d\overline{\sigma}^\text{ion}}{dW}\right) \!dW
\end{equation}
Note that the subscript $\mu$ here describes the mapping from an initial bin $[i]$ to \textit{all} possible set of final bins $\{[j],[k]\}$. Replacing the first integral by the usual Gaussian quadrature:
\begin{equation}
\label{eq:ion_zi2}
		\left. \frac{d \zebra_p (i) }{dt} \right]^{\text{ion}}_\mu = -N_l \frac{\varepsilon_{i 1} - \varepsilon_{i 0}}{2} \sum_{n[i ]} \omega_n U_p(\varepsilon_{i,n}) \, v_{i,n} f_i(\varepsilon_{i,n})\int_{\Delta E}^{\varepsilon_0}\left(\frac{d\overline{\sigma}^\text{ion}}{dW}\right) \!dW
\end{equation}
where $v_{i,n}\!=\!(2\varepsilon_{i,n}/m_e)^{1/2}$ and $\varepsilon_{i,n}$ is the value at a quadrature point of $[\varepsilon_{i0}, \varepsilon_{i1}]$. If we now discretize the values of $W$ according to the procedure \eqref{eq:4.41}:
\begin{equation}
\label{eq:ion_zi3}
		\left. \frac{d \zebra_p (i) }{dt} \right]^{\text{ion}}_\mu = -N_l \frac{\varepsilon_{i 1} - \varepsilon_{i 0}}{2} \sum_{n[i ]} \omega_n U_p(\varepsilon_{i,n}) \, v_{i,n} f_i(\varepsilon_{i,n}) \sum_{m=1}^{M} d \overline{\sigma}^{\text{ion}}_{\mu,n,m}
\end{equation}
where 
\begin{align}
d \overline{\sigma}^{\text{ion}}_{\mu,n,m} = \int_{W_{m-1}}^{W_m} \left(\frac{d\overline{\sigma}^\text{ion}}{dW}\right) \, dW
\end{align}
Finally, we perform the Legendre expansion for the distribution function and obtain:
\begin{equation}
\label{eq:ion_zi4}
		\left. \frac{d \zebra_p (i) }{dt} \right]^{\text{ion}}_\mu = - w_i N_l \sum_q \zebra_q (i) \sum_{n[i ]} \omega_n U_p(\varepsilon_{i,n}) U_q (\varepsilon_{i,n}) \, v_{i,n} \sum_{m=1}^{M} d \overline{\sigma}^{\text{ion}}_{\mu,n,m}
\end{equation}
Note that we have used here $\varepsilon_{i,n}$ as the energy of the incident electron. Since the quadrature points {$n$} are different for each mapping $\mu$, the discretized cross-section is also implicitly dependent on $\mu$; we have made this dependence more obvious by adding it to the indices of $d\overline{\sigma}^\text{ion}$. To rewrite eq. (\ref{eq:ion_zi4}) into a more compact form, let us define $k_{\mu,n,m}^\text{ion} = v_{i,n} d\overline{\sigma}^\text{ion}_{\mu,n,m}$ as the elementary rate of ionization. Finally, we obtain:
\begin{equation}\label{eq:ion_zi5}
\left. \frac{d\zebra_p(i)}{dt}  \right]^{\text{ion}}_\mu = -N_l\,\sum_q 
\zebra_q(i)\sum_{n[i]} \sum_m
 w_i\omega_n U_p(\varepsilon_{i,n})U_q(\varepsilon_{i,n}) k_{\mu,n,m}^\text{ion}
\end{equation}
Contrary to the simpler case of excitation/deexcitation, the final bin indices $j,k$ must be obtained for \emph{each} set of incident and transferred energy. 
Strictly speaking, the energy limits for the scattered electron are: $\varepsilon_{j0}=\varepsilon_{i,n}\!-\!W_m$, $\varepsilon_{j1}=\varepsilon_{i,n}\!-\!W_{m-1}$ while those of the ejected electron are $\varepsilon_{k0}=W_{m-1}\!-\!\Delta E$, and $\varepsilon_{k1}=W_m\!-\!\Delta E$. 
There is no guarantee that these limits fall within the same energy bin, and we would then have to consider all combinations. In the case where the energy grid is highly non-uniform with high resolution at lower energy, this leads to a very large number of possible combinations, and the cost associated with computing the rates becomes impractical. However, we will here simplify the problem by using the mid-point average $\hat{W}_{m}\!=\!(W_{m}\!+\!W_{m+1})/2$, in order to obtain a unique pair ($j,k$) of final bins (see Fig. \ref{fig:jump_ir}):
\begin{subequations}
\begin{align}\label{eq:4.52}
\varepsilon_{j,n} &= \varepsilon_{i,n}-\hat{W}_{m}\qquad \mapsto j\\
\varepsilon_{k,n} &= \hat{W}_{m}\!\!-\!\!\Delta E\qquad \mapsto k
\end{align}
\end{subequations}

The rate of change of the bin variables of $j$ and $k$ can then be written as:
\begin{equation}
\label{eq:ion_zj10}
\left. \frac{d\zebra_p(j)}{dt}  \right]^{\text{ion}}_\mu = N_l\,\sum_q 
\zebra_q(i)\sum_{n[i]} \widetilde{\sum_m}
 w_i\omega_n U_p(\varepsilon_{j,n})U_q(\varepsilon_{i,n}) k_{\mu,n,m}^\text{ion}
\end{equation}
where the summation $\widetilde{\sum_m}$ means a \emph{restricted} sum on values of $m$ such that $\varepsilon_{i,n} - \hat{W}_m\in [j]$. Similarly for bin $k$, we have:
\begin{equation}
\label{eq:ion_zk10}
\left. \frac{d\zebra_p(k)}{dt}  \right]^{\text{ion}}_\mu = N_l\,\sum_q 
\zebra_q(i)\sum_{n[i]} \widetilde{\widetilde{\sum_m}}
 w_i\omega_n U_p(\varepsilon_{k,n})U_q(\varepsilon_{i,n}) k_{\mu,n,m}^\text{ion}
\end{equation}
where the summation $\widetilde{\widetilde{\sum_m}}$ means a \emph{restricted} sum on values of $m$ such that $\hat{W}_m - \Delta E \in [k]$.

The recombination process is computed using the same mapping, i.e. the same energy bins are used. 
The elementary rate of recombination from two phase-space elements of width $d\varepsilon_1$ and $d\varepsilon_2$ -- starting at $\varepsilon_1$ and $\varepsilon_2$ -- into an infinitesimal bin $[\varepsilon_0, \varepsilon_0 + d\varepsilon_0]$ is (see eq. (\ref{eq:ir_f})):
\begin{equation}\label{eq:4.60}  
N_u^+ d\varepsilon_1 d\varepsilon_2 v_1 v_2 f(\varepsilon_1) f(\varepsilon_2) \sigma^\text{rec} \delta ( \{ \varepsilon \})
\end{equation}
Similarly, the double integral over $\varepsilon_1$ and $\varepsilon_2$ can be replaced by an integral over $W$. Using the same notation for all the bin indices $(i,j,k)$ ($j$ and $k$ are now the initial bins, and $i$ the final bin), the rate of change of the moment variables of bin $i$ is:
\begin{subequations}
\begin{align}
\left. \frac{d\zebra_p (i)}{dt} \right]^{\text{rec}}_\mu &= N_u^+ \int_{\varepsilon_{i 0}}^{\varepsilon_{i 1}} d\varepsilon_0 \,  U_p(\varepsilon_0) \int_{\Delta E}^{\varepsilon_0} v_1 \, v_2 \, f_j(\varepsilon_1) f_k(\varepsilon_2) \sigma^\text{rec} \, dW 
\end{align}
\end{subequations}
Following the same procedure as done in ionization, we replace the first integral by the Gaussian quadrature and the second integral by the summation over discrete energy transfer. We then introduce the Legendre expansions for $f_j$ and $f_k$ to obtain:
\begin{subequations}
\begin{align}
\label{eq:rec_zi10}
\left. \frac{d\zebra_p (i)}{dt} \right]^{\text{rec}}_\mu &= N_u^+ \frac{\varepsilon_{i1} - \varepsilon_{i0}}{2} \sum_{q,h}  \zebra_q(j) \, \zebra_h(k) \, \sum_{n[i]} {\sum_m} \, \frac{4}{\Delta_j \varepsilon \Delta_k \varepsilon}  \omega_n U_p(\varepsilon_{i,n})  U_q (\varepsilon_{j,n}) U_h (\varepsilon_{k,n}) \, v_{j,n} v_{k,n} \, \sigma^\text{rec} 
\end{align}
\end{subequations}
where $p,q,h$ are the polynomial indices. Defining $k^{\text{rec}}_{\mu,n,m} = v_{j,n} v_{k,n} \, \sigma^\text{rec}$ and rewriting the weight factor as $\frac{\varepsilon_{i1} - \varepsilon_{i0}}{2} \frac{4}{\Delta_j \varepsilon \Delta_k \varepsilon} = w_i \left[\frac{2\Delta_i\varepsilon}{\Delta_j\varepsilon \Delta_k\varepsilon}\right]$, we finally obtain:
\begin{subequations}
\begin{align}
\left. \frac{d\zebra_p (i)}{dt} \right]^{\text{rec}}_\mu &= N_u^+ \sum_{q,h}  \zebra_q(j) \, \zebra_h(k) \, \sum_{n[i]} {\sum_m} \, \left(\frac{2\Delta_i\varepsilon}{\Delta_j\varepsilon \Delta_k\varepsilon}\right)  \, w_i \omega_n  U_p(\varepsilon_{i,n})  U_q (\varepsilon_{j,n}) U_h (\varepsilon_{k,n}) k^{\text{rec}}_{\mu,n,m} 
\end{align}
\end{subequations}
The rates for the initial bins $(j,k)$ are can be derived similarly, leading to:
\begin{subequations}
\begin{align}
\label{eq:rec_zj10}
\left. \frac{d\zebra_p (j)}{dt} \right]^{\text{rec}}_\mu &= -N_u^+ \sum_{q,h}  \zebra_q(j) \, \zebra_h(k) \, \sum_{n[i]} \widetilde{\sum_m}  \left(\frac{2\Delta_i\varepsilon}{\Delta_j\varepsilon \Delta_k\varepsilon}\right)  \, w_i \omega_n  \,  U_p(\varepsilon_{j,n}) \,   U_q (\varepsilon_{j,n}) U_h (\varepsilon_{k,n}) \, k^{\text{rec}}_{\mu,n,m} \\
\label{eq:rec_zk10}
\left. \frac{d\zebra_p (k)}{dt} \right]^{\text{rec}}_\mu &= -N_u^+ \sum_{q,h}  \zebra_q(j) \, \zebra_h(k) \, \sum_{n[i]} \widetilde{\widetilde{\sum_m}}  \left(\frac{2\Delta_i\varepsilon}{\Delta_j\varepsilon \Delta_k\varepsilon}\right)  \, w_i \omega_n  \,  U_p(\varepsilon_{k,n}) \, U_q (\varepsilon_{j,n}) U_h (\varepsilon_{k,n}) \, k^{\text{rec}}_{\mu,n,m}
\end{align}
\end{subequations}
 Note that the summation over $m$ in (\ref{eq:rec_zj10}-\ref{eq:rec_zk10}) is restricted to values of the energy transfer consistent with the bins $(j,k)$, as for the ionization case (\ref{eq:ion_zj10}-\ref{eq:ion_zk10}). Similar to the case of ED processes, the elementary rates can be precomputed for all the mapping and IR processes. There is, however, an exception, that is, when we consider the effect of continuum lowering. In that case, the energy gaps can change as a function of time. However, it is reasonable to expect that the changes are gradual enough, such that only a small fraction of the mappings needs to be changed per time step. This will be investigated in the future.

\section{Elastic collisions/Thermalization}
\label{sec:thermalization}
In this section, we describe the discretization procedure for the elastic collision term in (\ref{eq:BFP}) and (\ref{eq:fluxee}). Let us recall that this term appears as a convection-diffusion equation in energy space:
\begin{align}
\label{eq:fp}
\partial_t f = - \partial_\varepsilon J_{ee}
\end{align}
The appropriate boundary condition for (\ref{eq:fp}) is that the flux $J_{ee}$ vanishes at $\varepsilon = 0$ and $\infty$. From here on, the subscript $ee$ is dropped for simplicity. To numerically solve this equation, we employ a discontinuous Galerkin (DG) discretization since it is consistent with the moment expansion introduced in the previous section. Multiplying eq. (\ref{eq:fp}) with the basic function $U_p$ and integrating over cell $i$, we obtain the following system of ODE's for the coefficients:
\begin{align}
\label{eq:dg}
\frac{\Delta_i \varepsilon}{2} \frac{d a_p^i}{dt} = - \left[ \hat{J}_{i+1/2} U_p (1) - \hat{J}_{i-1/2} U_p (-1) \right] + \int_{[i]} J \, \frac{dU_p}{d\varepsilon} \, d\varepsilon
\end{align}
where $\int_{[i]} \equiv \int_{i-1/2}^{i+1/2}$ and $\hat{J}$ denotes the numerical flux at the boundary of the cells. Note that eq.~(\ref{eq:dg}) can also be written in terms of the renormalized variable $\zebra_p (i)$ in a straightforward manner. The second term in (\ref{eq:dg}) represents an integral over the cell, which can be approximated using Gaussian quadratures:
\begin{align}
\int_{[i]} J \, \frac{dU_p}{d\varepsilon} \, d\varepsilon \simeq \frac{\Delta_i \varepsilon}{2}\sum_{n[i]} \omega_n J (\varepsilon_{i,n}) \frac{dU_p}{d\varepsilon} (\varepsilon_{i,n})
\end{align} 
However, as shown later in this section, we can rewrite this term into a flux form to better illustrate energy conservation. The flux can be split into a convective and diffusive part:
\begin{align}
\hat{J}_{i+1/2} = \hat{J}^C_{i+1/2} (f) + \hat{J}^D_{i+1/2} (f)
\end{align}
{From (\ref{eq:fluxee}), the convective and diffusive fluxes are:
\begin{subequations}
\begin{align}
\label{eq:Jconv}
\hat{J}^C_{i+1/2} &= \gamma \left( K_{i+1/2}/2\varepsilon_{i+1/2} - L_{i+1/2} \right) f^C_{i+1/2}\\
\label{eq:Jdiff}
\hat{J}^D_{i+1/2} &= -\gamma K_{i+1/2} \partial_\varepsilon (f^{D}_{i+1/2})
\end{align}
\end{subequations}
where $f^C_{i+1/2}$ and $f^D_{i+1/2}$ are the average values of $f$ at the boundary for convection and diffusion respectively.} 

The convective value is defined using the Chang-Cooper method\cite{chang_practical_1970} 
\begin{align}
f^C_{i+1/2} = \delta_{i+1/2} f^-_{i+1/2} + (1-\delta_{i+1/2}) f^+_{i+1/2}
\end{align}
Here $f^-_{i+1/2}$ denote the value of $f$ evaluated at the right boundary of cell $i$ and $f^+_{i+1/2}$ the value of $f$ evaluated at the left boundary of cell $i+1$. The fraction $\delta_{i+1/2}$ is defined as:
\begin{align}
\delta_{i+1/2} = \frac{1}{\varpi_{i+1/2}} - \frac{1}{\exp (\varpi_{i+1/2}) - 1}
\end{align}
where $\varpi_{i+1/2} = \frac{C_{i+1/2}}{D_{i+1/2}} (\varepsilon_{i+1} - \varepsilon_{i})$. The convective and diffusive coefficients, $C_{i+1/2}$ and $D_{i+1/2}$, can be easily derived from (\ref{eq:Jconv}) and (\ref{eq:Jdiff}). This type of average is commonly used in discretization of Fokker-Planck equation to preserve the correct steady-state solution \cite{epperlein_implicit_1994,kingham_implicit_2004,yoon_fokker-planck-landau_2014}. The original Chang-Cooper scheme was derived using a velocity (speed) grid; in this work, we use a version similar to Buet and Le Thanh\cite{buet_efficient_2008}.

For the diffusive term, the recovery-based DG scheme of van Leer and Nomura\cite{van_leer_discontinuous_2005} is employed. 
The method is briefly described here, while detail can be found in their paper. To obtain the average value $f^D_{i+1/2}$, we first recover a polynomial $g \equiv g (f_i, f_{i+1})$ that is continuous across two adjacent cells $i$ and $i+1/2$ from a $L_2$ minimization:
\begin{subequations}
\begin{align}
\int_{[i]} (g - f_i) \, U_p \, d\varepsilon &= 0\\
\int_{[i+1]} (g - f_{i+1}) \, U_p \, d\varepsilon &= 0
\end{align}
\end{subequations}
If $f$ is represented by a polynomial of order $p$, the recovered polynomial $g$ is of order $2p+1$. The boundary value and its first derivative can then be computed from the recovered polynomial, leading to a straightforward calculation of the diffusive flux.

Let us now examine the conservation properties of this scheme, which are not obvious due to the boundary conditions. The semi-discrete systems, written for the first two coefficients from eq.~(\ref{eq:dg}), are as follows:
\begin{subequations}
\begin{align}
\label{eq:dg0}
\frac{\Delta_i \varepsilon}{2} \frac{d a_0^i}{dt} &= - \left[ \hat{J}_{i+1/2} U_0 (1) - \hat{J}_{i-1/2} U_0 (-1) \right]\\
\label{eq:dg1}
\frac{\Delta_i \varepsilon}{2} \frac{d a_1^i}{dt} &= - \left[ \hat{J}_{i+1/2} U_1 (1) - \hat{J}_{i-1/2} U_1 (-1) \right] + \int_{[i]} J \, \frac{dU_1}{d\varepsilon} \, d\varepsilon
\end{align}
\end{subequations}
We are only interested in the first two coefficients, because number density and energy can be written as a linear combination of these two. In eq. (\ref{eq:dg0}), The integral over cell $i$ does not appear because $U_0$ is a constant. Since $\overline{n}_i = \frac{\Delta_i \varepsilon}{\sqrt{2}} a_0^i$, the rate of change of the total number density can be obtained by summing (\ref{eq:dg0})  over all the cells:
\begin{align}
\label{eq:dgden}
\int_0^\infty \partial_t f \, d\varepsilon \simeq 
\sum_{i=1}^{N_b} \frac{d\overline{n}_i}{dt} = \hat{J}_{1/2} - \hat{J}_{N_b+1/2}
\end{align}
Here the only two remaining terms are the fluxes at the left and right boundaries of the domain, which must be zero from the boundary conditions; therefore the total number density is conserved. For energy, using $\overline{e}_i = \overline{n}_i \varepsilon_i + \frac{(\Delta_i \varepsilon)^2}{2\sqrt{6}} a_1^i$ and the definition of $U_1$, we can show that the rate of change of the bin energy is simply:
\begin{align}
\label{eq:dgener}
\frac{d \overline{e}_i}{dt} = \int_{[i]} J \, d\varepsilon
\end{align}
where the flux terms canceled out exactly and only the integral remains. Inserting the definition of $J$ into (\ref{eq:dgener}) and integrating by parts, we obtain:
\begin{align}
\label{eq:dgener2}
\int_{[i]} J \, d\varepsilon = - \gamma \left[ Kf \right]^{i+1/2}_{i-1/2} + \gamma \int_{[i]} \left( \frac{K}{2\varepsilon} + \partial_\varepsilon K - L \right) f \, d\varepsilon
\end{align}
The first two terms inside the integrand can be grouped together as:
\begin{align}
\frac{K}{2\varepsilon} + \partial_\varepsilon K = \varepsilon^{-1/2} \partial_\varepsilon \left( \varepsilon^{1/2} K \right) = 3 \int_\varepsilon^\infty \varepsilon'^{-1/2} \, f \, d\varepsilon'
\end{align}
The last step is done by inserting the definition of $K$ and differentiating. Substituting the above equation back to (\ref{eq:dgener2}) and inserting the full expression of $L$ yield:
\begin{align}
\label{eq:dgener3}
\int_{[i]} J \, d\varepsilon = - \gamma \left[ Kf \right]^{i+1/2}_{i-1/2} + \gamma \int_{[i]} \left( 3f \int_\varepsilon^\infty \varepsilon'^{-1/2} \, f \, d\varepsilon' - 3 f \varepsilon^{-1/2} \int_0^\varepsilon f \, d\varepsilon' \right) \, d\varepsilon
\end{align}
The second term in the integrand can be integrated by parts, resulting into two terms, one of which cancels out the first term in the integrand exactly. We finally arrive at:
\begin{align}
\label{eq:dgener3b}
\int_{[i]} J \, d\varepsilon = - \gamma \left[ Kf \right]^{i+1/2}_{i-1/2} + \gamma \left[ \left( \int_\varepsilon^\infty 3 \varepsilon'^{-1/2} \, f \, d\varepsilon' \right) \left( \int_0^\varepsilon f \, d\varepsilon' \right) \right]^{i+1/2}_{i-1/2}
\end{align}
At this point, the integral over cell $i$ has been replaced by a net flux from the boundaries. Using this expression, the rate of change of the total energy becomes:
\begin{align}
\label{eq:dgener4}
\int_0^\infty \varepsilon \, \partial_t f \, d\varepsilon \simeq  \sum_{i=1}^{N_b} \frac{d\overline{e}_i}{dt} = \gamma \left( K_{1/2} f_{1/2} - K_{N_b+1/2} f_{N_b+1/2} \right)
\end{align}
In the limit of $N_b\rightarrow \infty$ and $\varepsilon_{N_b+1/2} \rightarrow \infty$, it is straightforward to see that the right hand size goes to zero, and the total energy is conserved. In the discrete case with finite truncation of the domain, eq. (\ref{eq:dgener4}) is zero iff $f_{N_b+1/2} = 0$. Note that a physical property of the distribution function $f$ is that it must decay to zero as $\varepsilon \rightarrow \infty$. This condition is not strictly enforced by the boundary condition so in the discrete case, energy conservation is not exact. Therefore, the DG scheme described so far is only mass conserving and not energy conserving. To obtain energy conservation, we propose to approximate eq. (\ref{eq:dgener3}) as follows:
\begin{align}
\label{eq:dgener5}
\int_{[i]} J \, d\varepsilon \simeq - \gamma \left[ K(f - f_{N_b+1/2}) \right]^{i+1/2}_{i-1/2} + \gamma \left[ \left( \int_\varepsilon^\infty 3 \varepsilon'^{-1/2} \, f \, d\varepsilon' \right) \left( \int_0^\varepsilon f \, d\varepsilon' \right) \right]^{i+1/2}_{i-1/2}
\end{align}
such that in the limit of $f_{N_b+1/2} \rightarrow 0$, we recover eq. (\ref{eq:dgener3}) exactly. This additional term acts as a correction to the energy flux to ensure conservation. 
Using (\ref{eq:dgener5}) in place of the integral in (\ref{eq:dg1}), the total energy in the discrete form is conserved exactly; this will be confirmed numerically in section {\ref{sec:numerics}}.

An interesting point to note is that for the special case of the $ee$ collision term and an uniform grid, it is possible to obtain energy conservation with a zeroth-order expansion ($N_p=0$) using the Rockwood's method\cite{rockwood_elastic_1973}. The essence of Rockwood's method is to alter the rates to satisfy energy conservation. In the paper of D'Angola et al. \cite{dangola_efficient_2010}, the authors introduced a fast method to evaluate these rates, and showed that it is more efficient compared to a standard method. It is unclear to us how this procedure is applied to the case of inelastic collisions. A comparison between Rockwood's method and ours is an interesting study for future work.

\section{Time integration}
Following the discretization described in the previous sections, we arrive at a semi-discrete system of ODE's both for the EEDF and ASDF:
\begin{align}
\label{eq:ode}
\frac{d\bm{y}}{dt} = \bm{R} (\bm{y})
\end{align}
where $\bm{y}$ denotes the vector of polynomial coefficients for all electron bins ($a_p^i$ or $\zebra_p (i)$) as well as atomic state densities ($N_k$). The numerical stiffness induced from a wide range of processes included in $\bm{R}$ suggests that an implicit time integration is preferred. A backward Euler method is applied to eq. (\ref{eq:ode}), yielding:
\begin{align}
\label{eq:ode2}
\frac{\Delta \bm{y}^n }{\Delta t} = \bm{R} (\bm{y}^{n+1})
\end{align}
where the superscript $n$ denote the current time step and $\Delta \bm{y}^n = \bm{y}^{n+1} - \bm{y}^n$. It must be pointed out that for our application, $\bm{R}$ is generally bi-linear in $\bm{y}$ except for the terms corresponding to the three body recombination process. To proceed, we first linearize eq. (\ref{eq:ode2}) about the current solution $\bm{y}^n$:
\begin{align}
\label{eq:ode3}
\bm{R} (\bm{y}^{n+1}) \simeq \bm{R} (\bm{y}^{n}) + \frac{\partial \bm{R}}{\partial \bm{y}} \Delta \bm{y}^n
\end{align}
where $\frac{\partial \bm{R}}{\partial \bm{y}}$ is the Jacobian matrix. Inserting (\ref{eq:ode3}) into (\ref{eq:ode2}), we obtain the following linear system to be solved at each time step:
\begin{align}
\label{eq:ode4}
\left[ \Delta t^{-1} \bm{I} - \frac{\partial \bm{R}}{\partial \bm{y}} \right] \times \Delta \bm{y}^n = \bm{R} (\bm{y}^n)
\end{align}
with $\bm{I}$ being the identity matrix. {Although the linearization procedure can introduce some errors in the energy conservation, numerical tests show that they are very small and essentially indistinguishable from numerical round-off errors. This will be illustrated in Sec \ref{sec:numerics}.} We also note that the Jacobian matrix is generally dense, with the exception of blocks connecting two ionization stages whose charges are differed by two or greater. The reason is that we only consider single step ionization kinetics, hence there is no process that directly results in doubly or triply ionization. 

{We point out that the time integration of Eq.~(\ref{eq:ode}) is independent of the discretization of the collision terms, hence other techniques can be used.} For example, one can consider an operator splitting approach, where the ASDF and the EEDF can be solved separately. However, some iterative procedure would be needed to achieve self-consistency among the solutions of the ASDF and EEDF. Furthermore, the time discretization of the $ee$ collision term alone could be done via an iterative procedure, similar to the approach of D'Angola et al.\cite{dangola_efficient_2010} and Epperlein\cite{epperlein_fokkerplanck_1994}, where the convective and diffusive coefficients are assumed to be slowly varying in time and can be iteratively updated within a time step. A comparison of the relative performance between the numerical approaches to time integration would depend on the details of implementation, size of the system to be solved, and the physical conditions of the problem; therefore, it remains beyond the scope of the present work, albeit an interesting study to be performed  in the future.

\section{Numerical Results}
\label{sec:numerics}
\subsection{Verification}
\label{sec:acc_tests}

\begin{figure}
	\centering
    \subfloat[]{
    		\includegraphics[scale=0.54]{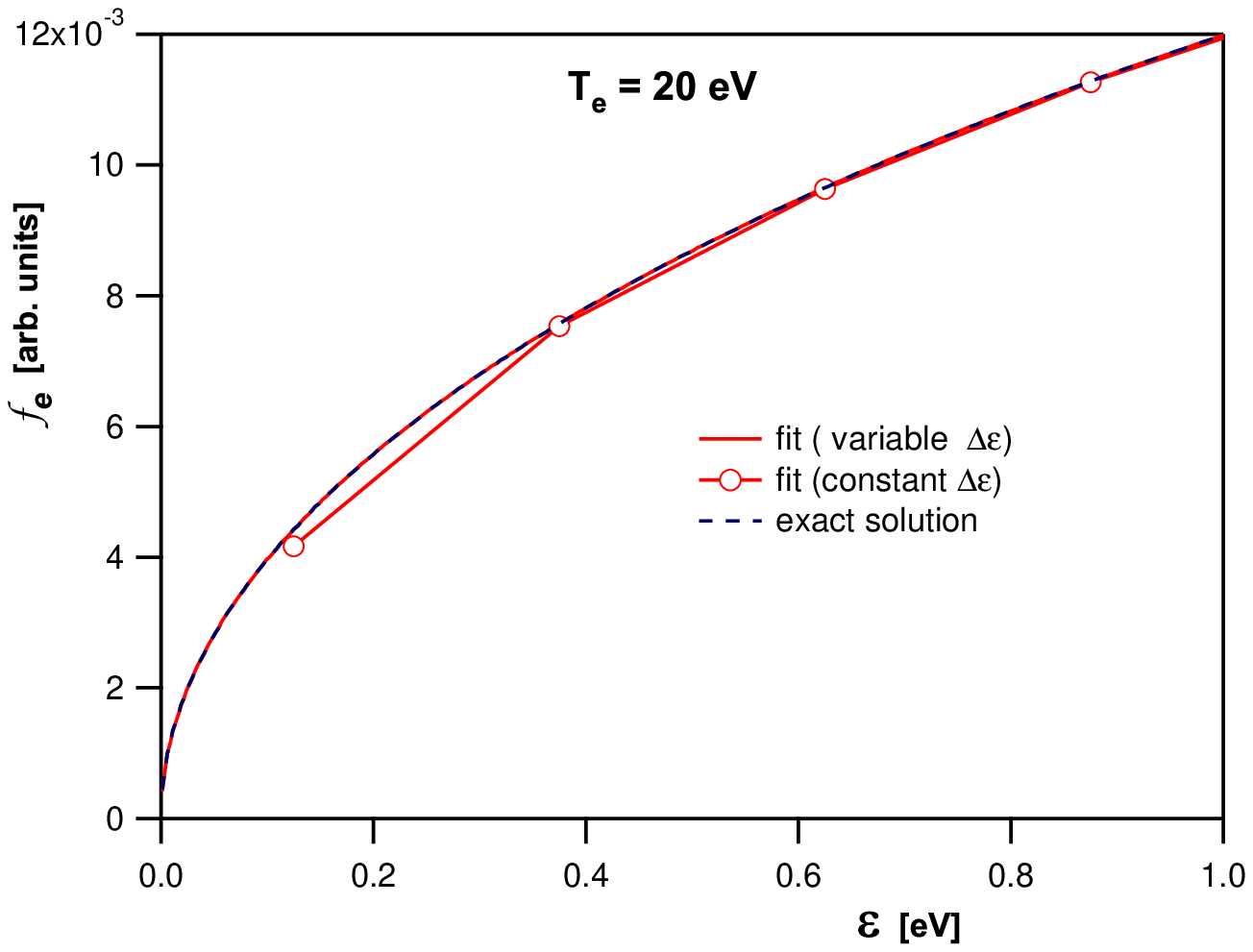}
        		\label{fig:comp_eedf1-a}
    }
    \subfloat[]{
    		\includegraphics[scale=0.54]{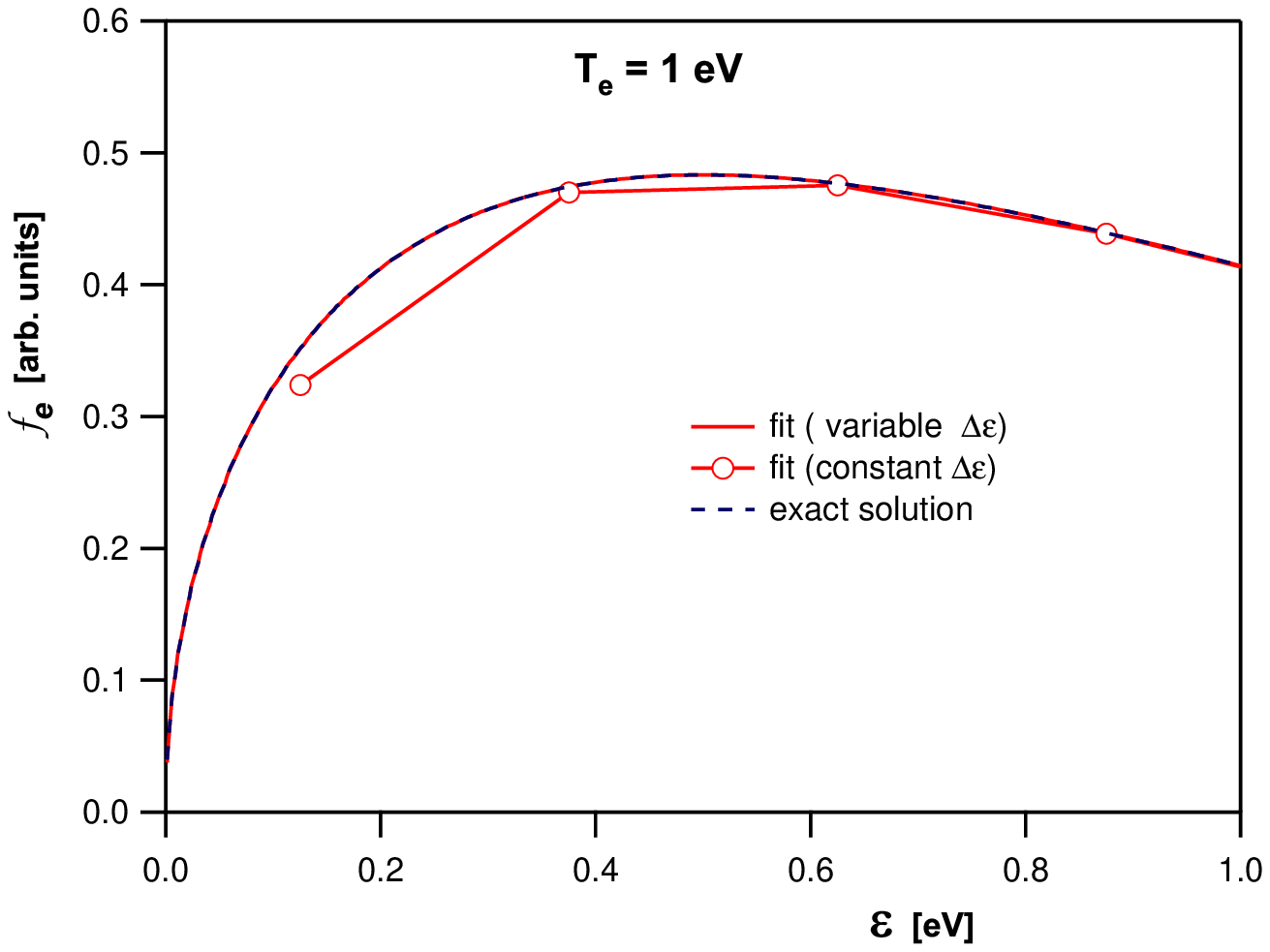}
     		\label{fig:comp_eedf1-b}
    }
    \caption{\footnotesize{Comparison of Maxwellian EEDF resolution at; (a) $T_e=20$ eV and (b) $T_e=1$ eV. }}
    \label{fig:comp_eedf1}
\end{figure}

The EEDF can be discretized on an energy grid with constant or variable spacings. In order to allow a large range of energy values, the constant-spacing in energy sacrifices accuracy in the low-energy region, where gradients of the EEDF can be large for the Maxwellian case. We compare two cases with the same maximum energy (250 eV) in Fig. \ref{fig:comp_eedf1}, for two Maxwellian distributions at high (20 eV) and low (1 eV) temperatures using 1000 bins.
The symbols show the bin centers in the constant-spacing case; the initial spacing in the variable binning is 2.5 milli-eV and 
increases geometrically. The EEDF value at the bin centers is not too far from the exact solution, but it is clear that a zeroth-order 
or even a first-order interpolation would be significantly in error, especially at low temperatures. Reducing the spacing to resolve the low-energy component either severely restricts the energy range or dramatically increases the number of bins and memory requirement. 

\begin{figure}
	\centering
	\includegraphics[scale=0.7]{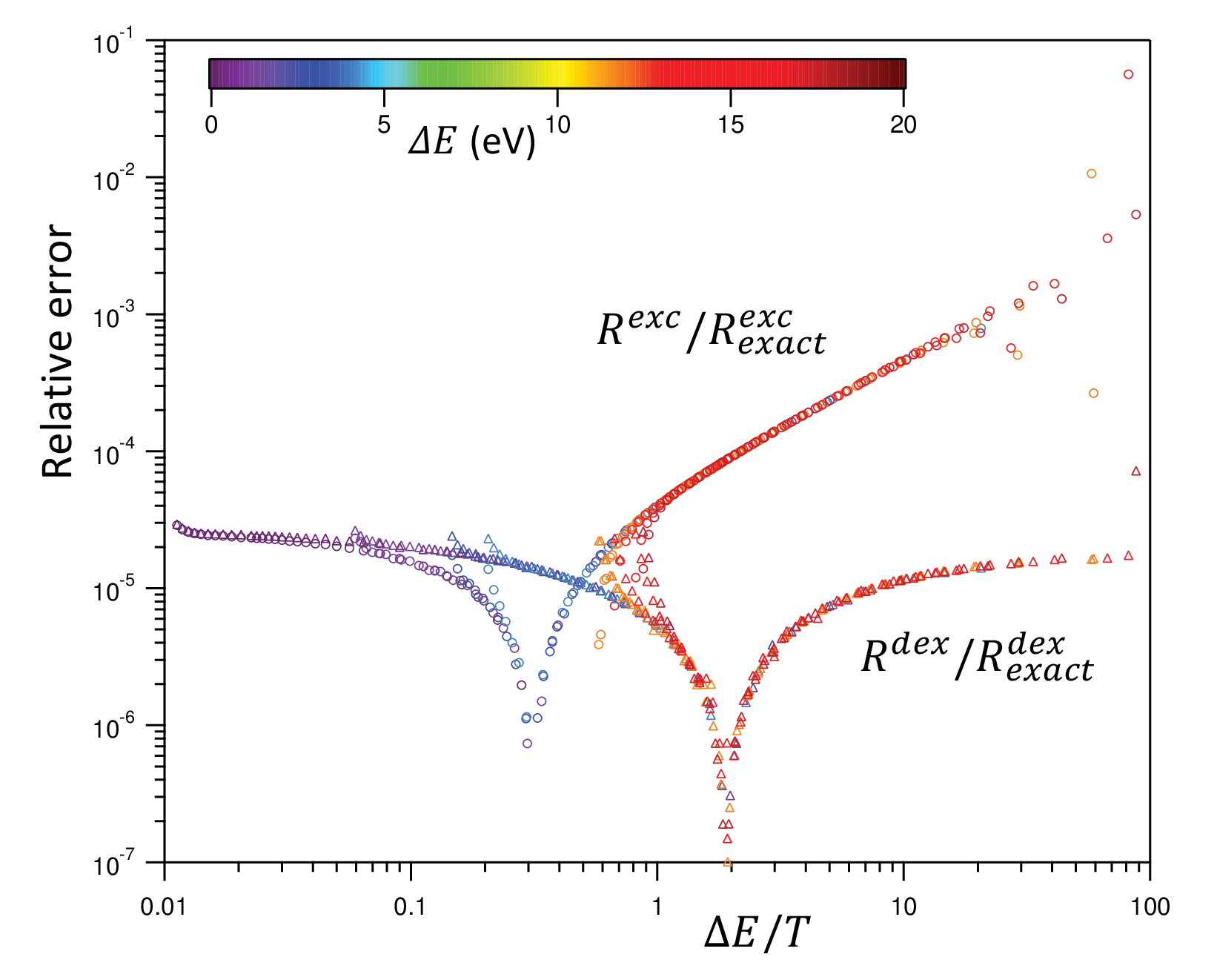}
	\caption{\footnotesize{Relative error of excitation ($R^{\text{exc}}$) and deexcitation ($R^{\text{dex}}$) rates for various transitions.}}
    \label{fig:rate_acc_xd}
\end{figure}

{The next step is to verify that the rates can be accurately computed with the discretized EEDF, using atomic Hydrogen as an example. Utilizing the classical form of the inelastic cross-sections, we can directly compare the rates with the analytical solution.} Fig. \ref{fig:rate_acc_xd} shows the relative error of the excitation and deexcitation rates as function of the normalized energy gap $\Delta E/T$. Various values of energy spacings are used, as indicated by the color spectrum of the symbols. The largest errors are obtained for large energy gaps at low temperatures, due to the exponentially small value of the excitation rate. Since the absolute rate at such conditions is very small and the excitation is essentially non-existent, this regime is of little consequence. In the regions of interest, the rates are computed with an accuracy better than $10^{-4}$. {The same can be done with ionization and recombination rates, which shows approximately the same level of accuracy. It is important to note that these discretization errors do not affect the energy conservation property of the method, because the formulation of the rate equations (using these approximate rates) always results in zero rate of change of total energy.}

\begin{figure}
\begin{center}
    \includegraphics[scale=.7]{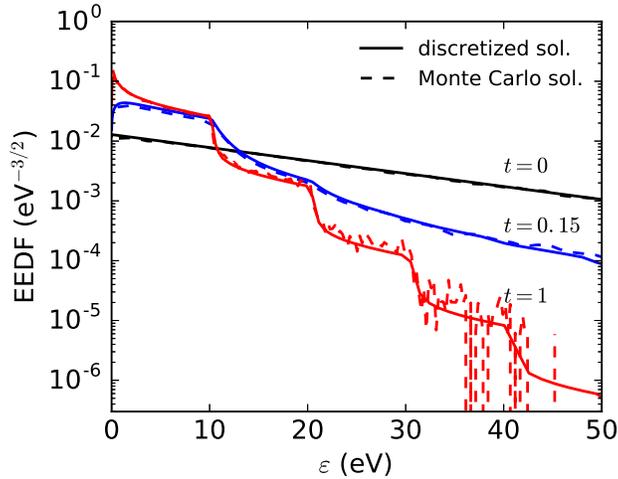}
\caption{\footnotesize{Numerical solution of the normalized EEDF at 0, 0.15 and 1 $\mu$sec. The dashed line is the result from a Monte Carlo calculation.}}
\label{fig:tla1}
\end{center}
\end{figure}

{In the next test, we simulate a model problem of a two level atom. This problem was first introduced in Yan et al.\cite{yan_analysis_2015} to demonstrate that for system with only excitation/deexcitation, the stationary solution of the EEDF is not a Maxwellian. This is due to the fact that electrons can only gain or lose discrete values of energy during a solely excitation/deexcitation process. 
A theoretical analysis of this system can be found in Yan et al.\cite{yan_analysis_2015} We consider here a Hydrogen atom with two states: a ground and an excited state. Only excitation/deexcitation is included, and the electron density stays constant. The initial atom density is set at $10^{15}$ cm$^{-3}$, and all particles are at the ground state. The initial condition for the EEDF is a Maxwellian at 20 eV with a density of $10^{14}$ cm$^{-3}$. The EEDF is discretized by 160 bins with a variable spacing starting at $\Delta \varepsilon_\text{min} = 10^{-2}$ eV, up to $\Delta \varepsilon_\text{max} = 2.5~10^{+2}$ eV. As the simulation evolves, the excited state begins to populate due to collision with the electrons. This simulation also demonstrates the advantage of a non-uniform grid, since the extent of the EEDF is greatly reduced as the electron cools down. A constant spacing grid with proper resolution at high energy, although needed in the beginning to capture the equilibrium EEDF at 20 eV, becomes wasteful at the end. Fig. \ref{fig:tla1} shows the numerical solution of the normalized EEDF, defined as $f/ N_e \sqrt{\varepsilon}$, at three instances of time: 0, 0.15 and 1 $\mu$sec, and the comparison with a similar calculation using the Monte Carlo collision algorithm.\cite{yan_analysis_2015} It can be seen that the EEDF develops discontinuities occurring at multiples of the energy gap of the excitation transition (10.2 eV). These features are captured both in the discretized and Monte Carlo solutions. Both solutions agree very well, confirming that the physics is correctly implemented. We also point out that while the discretized solution can resolve the EEDF down to $10^{-6}$ and below, the Monte Carlo method suffers from the numerical noise. Furthermore, the implicit time integration allows a large time step, making the algorithm suitable for examining systems at a long time scale.}

\begin{figure}
\begin{center}
\includegraphics[scale=0.7]{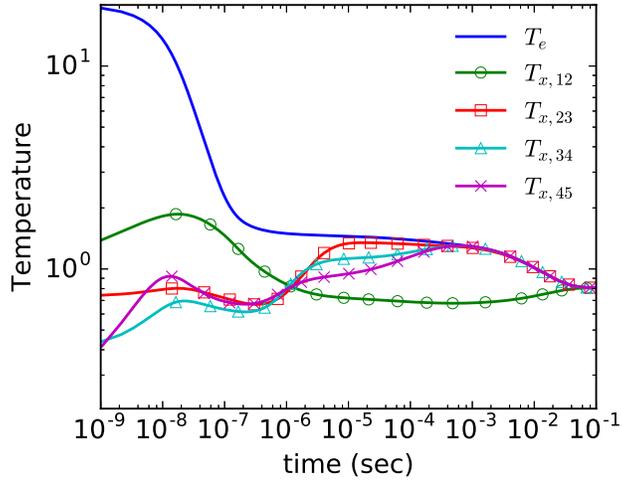}
\caption{\footnotesize{Temperature profiles as function of time. The Boltzmann temperatures between two levels are defined according to eq. (\ref{eq:Boltz}).}}
\label{fig:hydrogen2_Te}
\end{center}
\end{figure}

\begin{figure}
\begin{center}
\includegraphics[scale=0.7]{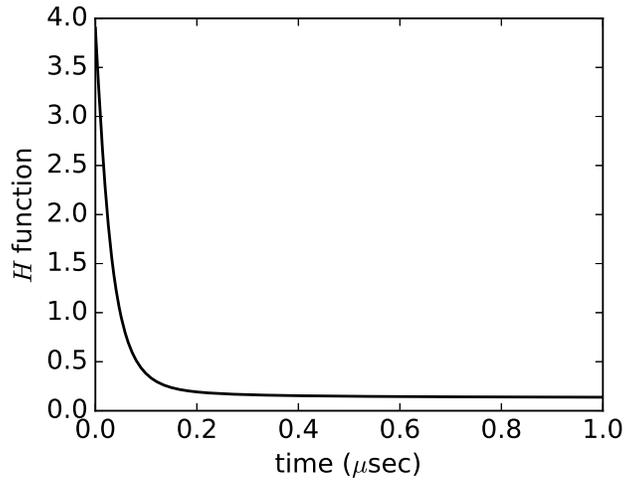}
\caption{\footnotesize{Time evolution of the total $H$ function (electron + atom) during the relaxation.}}
\label{fig:hydrogen2_Hfunc}
\end{center}
\end{figure}

\begin{figure}
\begin{center}
\includegraphics[scale=0.7]{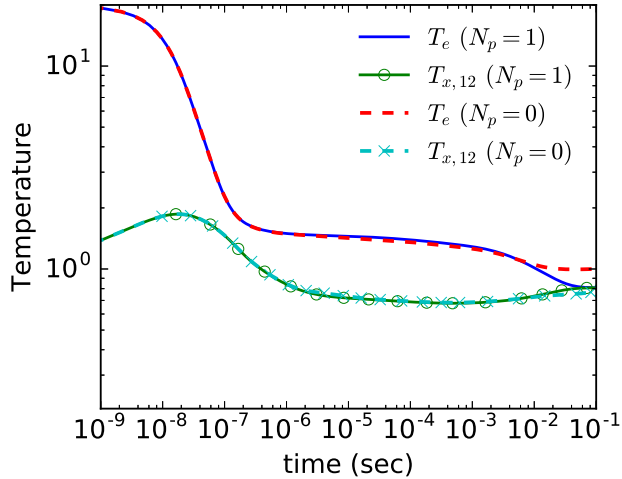}
\caption{\footnotesize{Temperature profile as a function of time. The solid lines show the solution with first-order and the dashed lines zeroth-order expansion.}}
\label{fig:hydrogen2_Tx}
\end{center}
\end{figure}

\begin{figure}
\begin{center}
\includegraphics[scale=0.7]{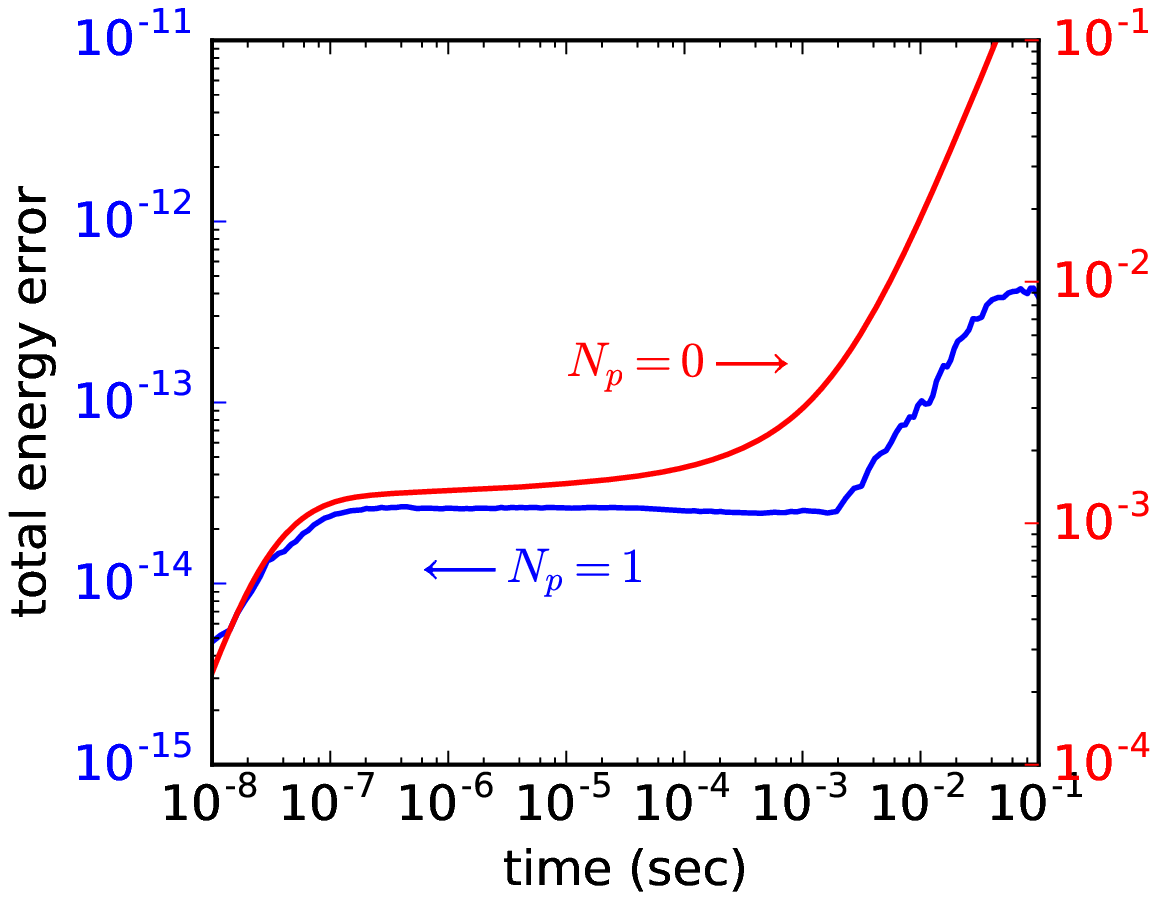}
\caption{\footnotesize{Relative error on total energy as function of time; left axis - first-order expansion; right axis - zeroth-order expansion.}}
\label{fig:hydrogen2_errtot}
\end{center}
\end{figure}

{Let us now demonstrate the conservation properties of the algorithms for inelastic collisions. We first note that for all the test cases shown here, mass (or charge) is always conserved, and numerical tests showed a relative accuracy of the order of $10^{-15}$ throughout the evolution of a non-equilibrium plasma. Energy conservation, however, is only achieved when the moment expansion (\ref{eq:expansion}) is carried to at least first-order ($N_p=1$). In this test case, we also consider an atomic Hydrogen plasma but with 5 atomic levels. The initial condition and discretization of the EEDF are similar to the previous test. At $t=0$, all neutral atoms are at the ground state, and the ion density is the equal to the electron density. Both excitation/deexcitation and ionization/recombination are included in the calculation. Fig. \ref{fig:hydrogen2_Te} shows the time evolution of the effective electron temperature (2/3 of the electron mean energy), and the Boltzmann temperatures between two adjacent atomic levels, defined according to eq. (\ref{eq:Boltz}) using the nonequilibrium values of the atomic densities. Due to the small energy gaps between them, the highly excited states tend to equilibrate quicker than the ground state (note the slow equilibration of $T_{x,12}$). It is clear that the system approaches the correct thermodynamic equilibrium limit at 0.8 eV. Fig. \ref{fig:hydrogen2_Hfunc} shows that the total $H$ function (negative of entropy) decreases monotonically, which is consistent with the H-theorem (see appendix \ref{app:h-theorem}). The $H$ function is defined similarly to Yan et al.\cite{yan_analysis_2015} (see eq. (3.13) in their work). We emphasize that energy conservation is crucial in achieving the correct equilibrium limit. To demonstrate the importance of energy conservation, Fig. \ref{fig:hydrogen2_Tx} shows a comparison of $T_e$ and $T_{x,12}$ for the two cases of zeroth and first-order expansions; the non-conservation of energy in the former case (shown in dashed lines) leads to erroneous values of the temperature, and the system does not approach equilibrium. This issue is further illustrated in Fig. \ref{fig:hydrogen2_errtot}, which shows the relative error in total energy as the plasma relaxes towards equilibrium. Note that the error in the zeroth-order expansion case (red line) is shown on the \emph{right} axis, and continuously increases during the plasma relaxation. By contrast, the error for the first-order expansion, with scale shown on the \emph{left} axis, always remains low values, many orders of magnitude below the zeroth-order case. The fact that the error growth is unchecked during the time evolution is the most damaging consequence of the non-conservative zeroth-order algorithm, i.e. we could not guarantee accuracy on the energy of the system unless special measures are taken. 
Although not shown here, the error of the total energy per time step indicates that energy is conserved down to round-off errors for the case of first-order expansion. }

\begin{figure}
    \centering
    \includegraphics[scale=.7]{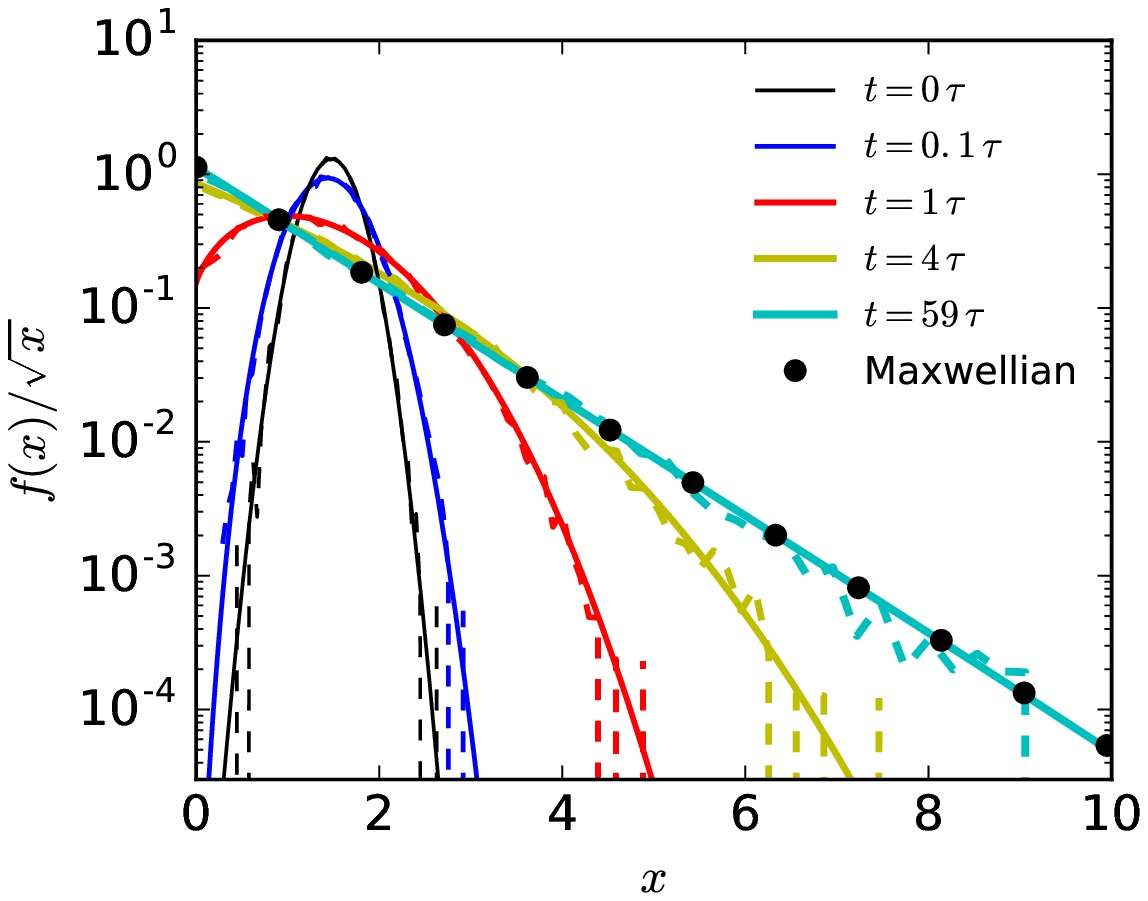}
    \caption{\footnotesize Time evolution of the normalized EEDF during a relaxation from an initial Gaussian distribution. Energy is normalized by the effective temperature $T$, and time is normalized by the collision time $\tau$. The numerical solutions from the Monte Carlo calculation are shown in dashed lines, and the circles show the analytical solution of a Maxwellian distribution.}
    \label{fig:thermalization}
\end{figure}

\begin{figure}
        \includegraphics[scale=.7]{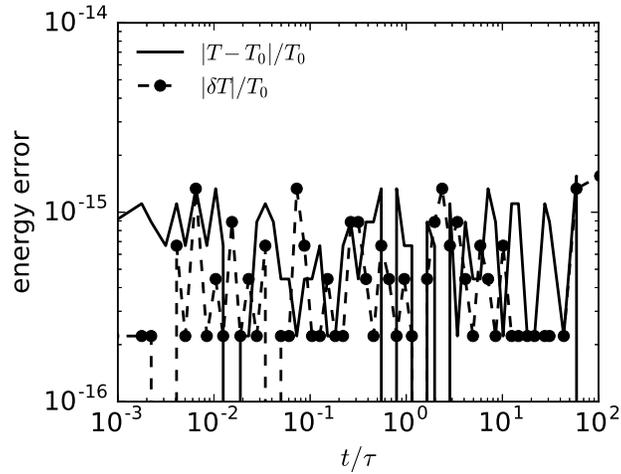}
    \caption{\footnotesize Relative error on total energy (solid) and error per time step (dashed with circles) during the relaxation shown in Fig. \ref{fig:thermalization}. Here $T_0$ refers to the effective temperature at $t=0$ and $\delta T$ to the change in the effective temperature per time step.}
    \label{fig:thermalization2}
\end{figure}

We now turn the attention to the $ee$ collision operator. In this next test case, we simulate the relaxation of an initially Gaussian EEDF. This is somewhat similar to the Rosenbluth problem commonly known in the literature \cite{rosenbluth_fokker-planck_1957}. Here only the $ee$ collision term is active, and the EEDF is expected to relax toward a Maxwellian distribution with the same initial total density $N_0$ and energy $E_0$. For this problem, we can define a non-dimensional energy variable as $x \equiv \varepsilon/T$ where $T=\frac{2}{3} E_0/N_0$, and a non-dimensional time $\hat{t} \equiv t/\tau$ where $\tau = m_e^2 v_t^3/\left[ 4\pi N_0 e^4 \ln \varLambda \right]$ and $v_t = \sqrt{T/m_e} $. The solution of the FP equation becomes self-similar under this transformation, and the steady-state limit is given by a non-dimensional Maxwellian EEDF, i.e., $f(x)=2 \pi^{-1/2} x^{1/2} e^{-x}$. In this test case, the initial Gaussian distribution has a mean and variance of $1.5$ and $0.25$, respectively. The EEDF is discretized by 100 bins and the grid spacings are chosen with higher resolution at lower energy. Fig. \ref{fig:thermalization} shows the time evolution of the normalized EEDF, which clearly relaxes toward a Maxwellian distribution as time evolves. At $t=59 \, \tau$, the EEDF is completely thermalized, matching the analytical Maxwellian distribution shown in the circles.  In order to check the accuracy of the algorithm, we also compare the solutions to those obtained from a Monte Carlo collision model\cite{takizuka_binary_1977}, the results of which are shown in dashed lines. Both models agree very well, albeit with some numerical noise in the Monte Carlo calculation.
To numerically confirm the energy conservation properties of the algorithm, Fig. \ref{fig:thermalization2} shows the relative error on total energy as the simulation progresses. Both the total error (solid) and error per time step (dashed) clearly indicate that energy is conserved down to round-off error. 

\subsection{Transient Heating of an Argon Cluster Target}
In this section, we apply the numerical algorithms developed in this work to a more complex system, consisting of many ions and atomic levels. We simulate a transient heating process of an Argon cluster target due to high energy electrons produced from fast laser absorption. We note that this test case is very challenging for the Monte Carlo algorithm due to the large number of processes involved. The numerical set-up is very similar to the work of Abdallah et al. \cite{abdallah_time-dependent_2003} We assume that initially all the atoms are in the ground state of Ne-like ion stage (Ar$^{8+}$), which approximates the plasma condition after the laser pre-pulse. The atomic data used in this test case includes all the ion stages from Ne-like to fully-stripped (Ar$^{18+}$), and is constructed based on the screened-hydrogenic formalism\cite{scott_advances_2010,chung_flychk:_2005}. All the energy levels are truncated at a maximum principle quantum number of 10, which is sufficient for most applications of interest. The atomic model consists of 187 atomic levels and approximately 1000 transitions including both excitation and ionization. The initial density of the atoms is $5.0~10^{20}$ cm$^{3}$. From charge neutrality, the electron density is $4.0~10^{21}$ cm$^{3}$. The EEDF is discretized into 100 non-uniform bins with finer resolution at low energy. The initial condition of the EEDF is a Gaussian with a mean and variance of 5 and 0.4 keV respectively. This corresponds to an effective temperature of 3.3 keV.

\begin{figure}
	\centering
    \subfloat[]{
    	\includegraphics[scale=.7]{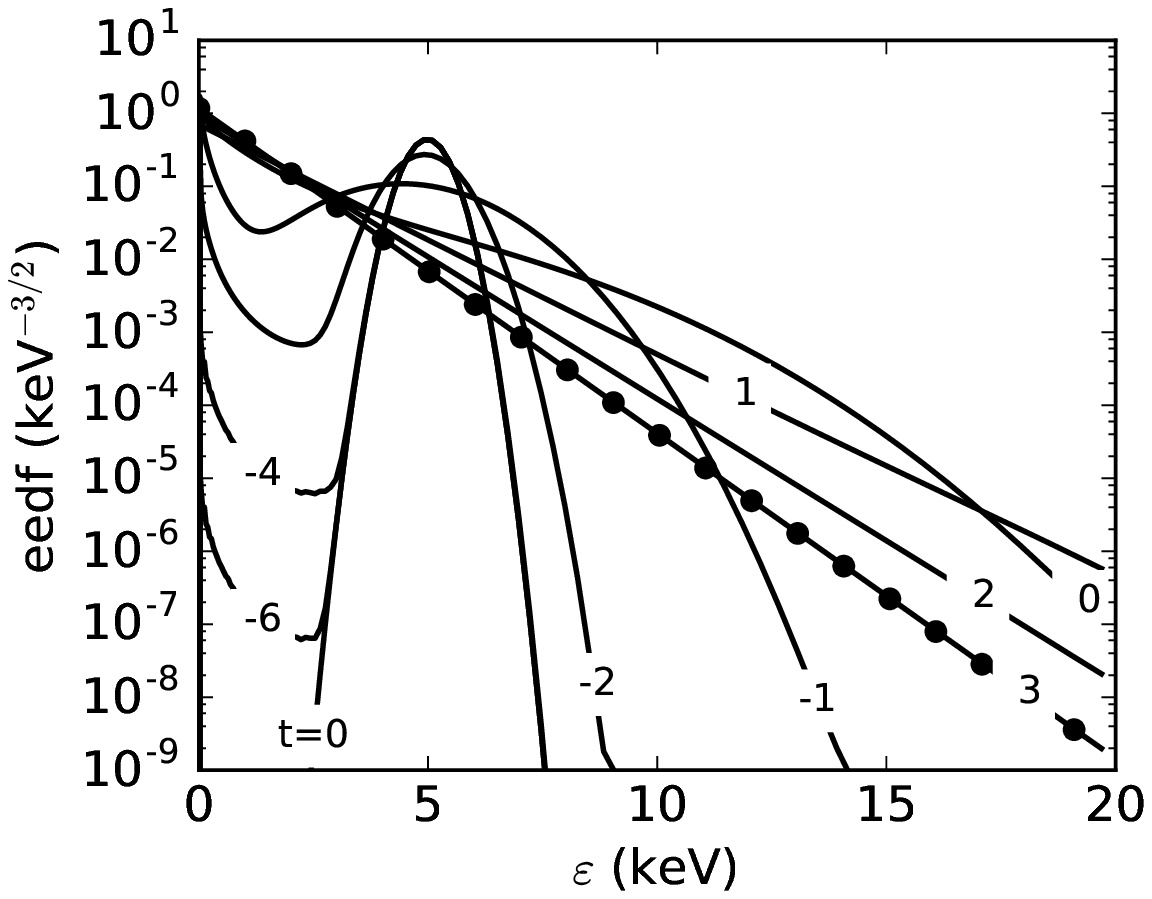}
    }
    \subfloat[]{
    	\includegraphics[scale=.7]{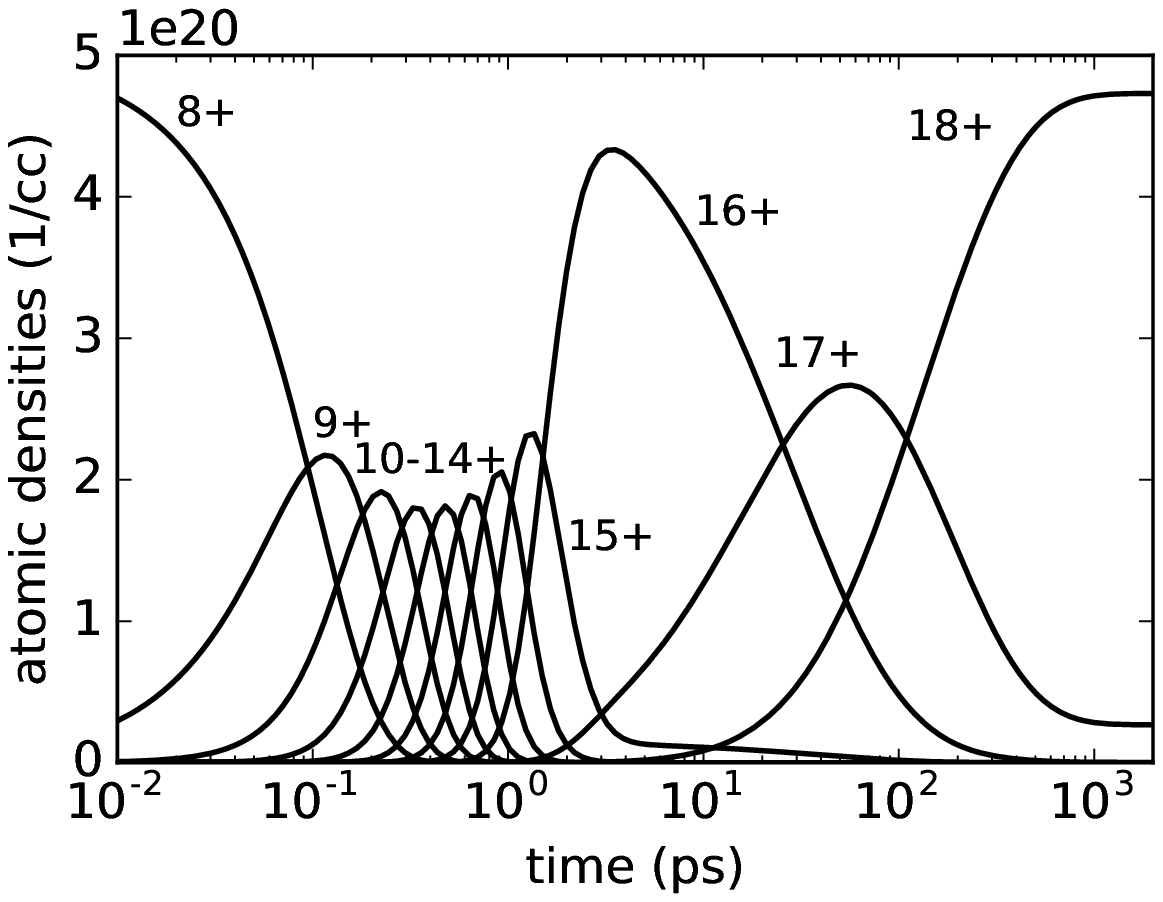}
    }
    \caption{\footnotesize Time evolution of (a) the normalized EEDF and (b) the ion densities from Ne-like (8+) to fully-stripped (18+). The curve labels in (a) (except for $t=0$) correspond to the exponent (with base 10) of times in ps.}
    \label{fig:arcluster}
\end{figure}

\begin{figure}
    \centering
        \includegraphics[scale=.7]{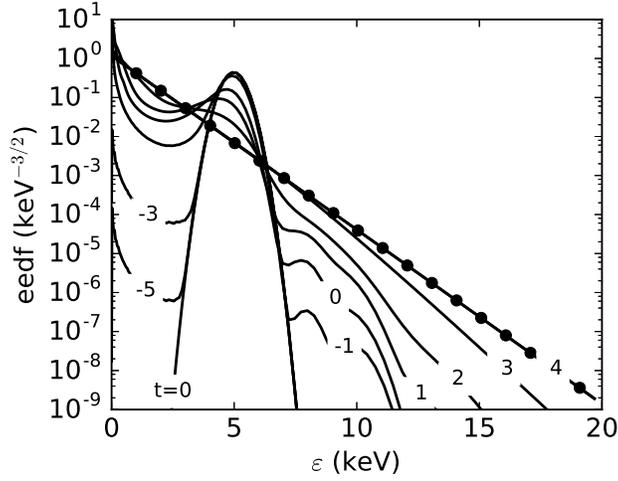}
    \caption{\footnotesize Time evolution of the normalized EEDF for the same test case in Fig. \ref{fig:arcluster} but without $ee$ collisions.}
    \label{fig:arcluster_noee}
\end{figure}

\begin{figure}
  \centering
    \includegraphics[scale=.7]{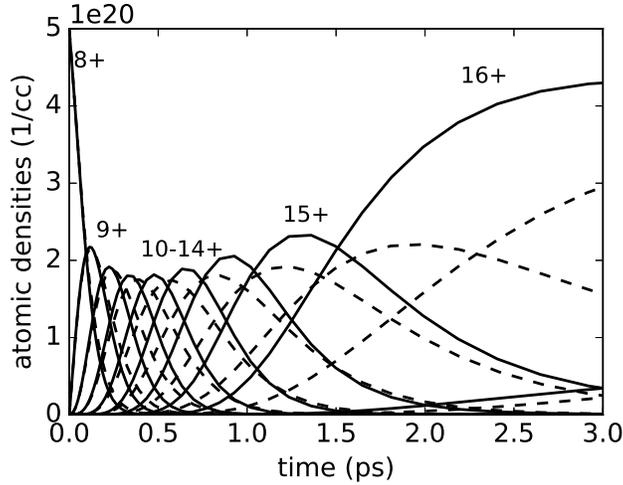}
    \caption{\footnotesize Time evolution of the ion densities with (solid) and without (dashed) $ee$ collisions.}
    \label{fig:arcluster2}
\end{figure}

{The time evolution of the normalized EEDF and the atomic density of each ion stage, i.e., summed over all the atomic levels, is shown in Fig. \ref{fig:arcluster}. It can be seen that the non-thermal electrons create a rapid and transient ionization with multiple ions appearing and disappearing from Ne-like sequence to the fully-striped nucleus. At 2 ns, both the EEDF and the ASDF are in thermodynamic equilibrium, i.e., Maxwellian and Boltzmann-Saha, and the EEDF matches the analytical distribution at the equilibrium temperature shown in circles. Fig. \ref{fig:arcluster}a also indicates that by 10 ps the EEDF has been completely thermalized, which is primarily due to $ee$ collisions. We also repeat the same test case but without $ee$ collisions, and the results show that the EEDF does not get to a Maxwellian shape until the very last time step (see Fig. \ref{fig:arcluster_noee}). This also leads to a significant difference in the time evolution of the ion densities as can be seen in Fig. \ref{fig:arcluster2}. For the case with $ee$ collisions, the ionization proceeds faster due to the formation of a high energy tail in the EEDF starting from 1 ps as shown in Fig. \ref{fig:arcluster}a. 
Lastly, we note that the energy error in this simulation also indicates that total energy is conserved down to round-off error.} 

\section{Conclusions}
\label{sec:conclusion}
In the current work, we introduced a set of algorithms for modeling non-Maxwellian plasma kinetics {including both inelastic and elastic collisions}. The new algorithms, based on an expansion of the EEDF in Legendre polynomials, are applied to the case of a non-uniform grid in energy space. By design, the proposed method automatically conserves density and energy exactly in each time step. We described the general procedure to construct source terms for the solution variables in the cases of excitation, ionization and electron-electron elastic collisions. The reverse processes of deexcitation and recombination are also included and their rates are computed based on the principle of detailed balance. The proposed algorithms are very computationally efficient, because all of the transition rates (among the EEDF bins) are pre-computed. The elastic collision term is discretized using a discontinuous Galerkin formulation. We propose to augment the standard discontinuous Galerkin source terms by an equivalent flux form to achieve exact energy conservation. The resultant discrete system is solved by a backward Euler time discretization, which alleviates the severe time step restriction due to numerical stability. Several numerical test cases are presented to demonstrate the capability of the method. We show that total energy is always conserved down to round-off errors, and the numerical solutions always approach the correct thermal equilibrium limit at large time. We successfully applied the current method to model a complex atomic system, consisting of many atomic levels and transitions. 

The present work can be extended in multiple directions. The radiative coupling terms, e.g., bound-bound, bound-free and free-free, can be incorporated in a similar manner, and the radiation transport equation can be solved self-consistently using the same discretization procedure. For very high density and low temperature plasmas, the effects of continuum lowering and electron degeneracy become significant and must be taken into account. Depending on the specific implementation of a continuum lowering model, some fraction of the transitions might need to be recomputed, and an optimization strategy can be utilized to speed-up this process. Incorporating the electron degeneracy requires a generalization of the collision operators, which introduces additional terms in the rate expression, e.g., Pauli blocking factors, but the discretization remains unchanged. {Furthermore, the efficiency of the current solver can be improved by exploring different numerical approaches to the solution of the discretized equations for the EEDF and ASDF}. All of these will be considered in future work.

\section*{Acknowledgments}
The authors would like to thank Professor Russel Caflisch and Dr. Howard Scott for many fruitful discussions. This work was performed under the auspices of the U.S. Department of Energy by Lawrence Livermore National Laboratory under Contract No. DE-AC52-07NA27344. HPL was partially supported by the Air Force Office of Scientific Research (AFOSR) grant FA9550-14-1-0283.

\numberwithin{equation}{section}
\begin{appendices}
\renewcommand\appendixname{Appendix}

 \section{H-theorem}
 \label{app:h-theorem}

For an ensemble of $\mathcal{N}$ particles, the (classical) number of independent micro-states available and compatible with a given total energy is:
\begin{equation}\label{eq:a2.21}
\mathcal{W} = \frac{\mathcal{G}^\mathcal{N}}{\mathcal{N}!}
\end{equation}
where $\mathcal{G}$ is the degeneracy of the particles. The entropy is given by (see sections 5.1.1 and 5.1.9 of Oxenius\cite{oxenius_kinetic_1986}):
\begin{equation}\label{eq:a2.22}
\mathcal{S}=k\ln\mathbf{\mathcal{W}}
\end{equation}
where $k$ is the Boltzmann constant. Using the Stirling formula $\ln(\mathcal{N}!)\approx \mathcal{N}\ln \mathcal{N}-\mathcal{N}$:
\begin{equation}\label{eq:a2.23}
\mathcal{S}\approx k\left[\mathcal{N}\ln \mathcal{G} -\mathcal{N}\ln \mathcal{N} + \mathcal{N}\right]=-k\mathcal{N}\left[\ln\left(\frac{\mathcal{N}}{\mathcal{G}}\right)-1\right]
\end{equation}
The number of particles in an energy state $\varepsilon$ is given by:
\begin{equation}\label{eq:a2.1}
\mathcal{N} (\varepsilon) d\varepsilon = \eta (\varepsilon) \mathcal{G} (\varepsilon) d \varepsilon
\end{equation}
where $\eta (\varepsilon)$ is the mean occupation number the quantum state. Following Oxenius\cite{oxenius_kinetic_1986} (pp. 10-13), the total degeneracy for electrons is:
\begin{equation}\label{eq:a2.2}
\mathcal{G} (\varepsilon) = \mathcal{V} g_e\, \underbrace{\frac{2^{5/2}\pi m_e^{3/2}}{h^3}\varepsilon^{1/2}}_{\mathcal{G}_\text{trans}(\varepsilon)} 
\end{equation}
where $\mathcal{V}$ is the volume, while the degeneracy of the atom or ion is:
\begin{equation}\label{eq:a2.3}
\mathcal{G} (A_l) = \mathcal{V} g_l \mathcal{G}_\text{trans}(\varepsilon_l)
\end{equation}
Strictly speaking, the degeneracy due to the translational degrees of freedom for the heavy particles should be included; however, the large mass difference between electrons and atoms 
implies that the center-of-mass is approximately co-located with the atom and ion, whose kinetic energy changes little. 
Therefore, one can ignore these (approximately constant) translational degrees of freedom in the formulation which follows.

Differentiating eq. (\ref{eq:a2.23}), the time-derivative of the entropy for particles of a given type is:
\begin{equation}\label{eq:a2.24}
\frac{\partial\mathcal{S}}{\partial t} = -k\left(\frac{\partial \mathcal{N}}{\partial t}\right) \left[\ln\left(\frac{\mathcal{N}}{\mathcal{G}}\right)-1\right]-k\left(\frac{\partial \mathcal{N}}{\partial t}\right)
\end{equation}
Using the occupation number, the time rate of change of the entropy can be written as:
\begin{equation}\label{eq:a2.25}
\frac{\partial\mathcal{S}}{\partial t} = -k\left(\frac{\partial \mathcal{N}}{\partial t}\right) \ln(\eta)
\end{equation}
Equation (\ref{eq:a2.25}) shows that the only time variation of the entropy is contained in the number of particles, weighted by the logarithm of the occupation number. One can also define variables per unit volume, i.e., total density $N=\mathcal{N}/\mathcal{V}$, differential density $N (\varepsilon)=\mathcal{N(\varepsilon)}/\mathcal{V}$ and entropy $S = \mathcal{S}/\mathcal{V}$. Note that $N(\varepsilon) d\varepsilon \equiv f(\varepsilon) d\varepsilon$.

Consider now the change in entropy from the collisional ionization and the reverse process, the entropy production rate can be written as:
\begin{eqnarray}\label{eq:a2.26}
\frac{\partial }{\partial t} \delta S&= -k & \left[\frac{\partial N(\varepsilon_0)}{\partial t}\ln\eta(\varepsilon_0)
                                                                    +\frac{\partial N (\varepsilon_1)}{\partial t}\ln\eta(\varepsilon_1)\right. \\
                                            &             &\left.+\frac{\partial N (\varepsilon_2)}{\partial t}\ln\eta(\varepsilon_2)
                                                                    +\frac{\partial N^+_u}{\partial t}\ln\eta_u^+ + \frac{\partial N_l}{\partial t}\ln\eta_l\right]\nonumber
\end{eqnarray}
{Note that here we consider a differential entropy $\delta S$; the rate of change of the total entropy of the system is simply obtained by integrating over all compatible electron energies, i.e., $\frac{\partial S}{\partial t} = \int \frac{\partial }{\partial t} \delta S$}. The rates of change of the particle densities can be obtained from the kinetic rates; using mass conservation and simplifying the notation,
\[
\frac{\partial  N_l }{\partial t} = \frac{\partial N (\varepsilon_0)}{d\partial } = - \frac{\partial N (\varepsilon_1)}{\partial t} = - \frac{\partial N (\varepsilon_2)}{\partial t} = - \frac{\partial N^+_u }{\partial t} = dR^\text{rec} - dR^\text{ion}
\]
yielding
\begin{equation}\label{eq:a2.27}
\frac{\partial }{\partial t} \delta S = k \,\left[dR^\text{rec}-dR^\text{ion}\right]\cdot\ln\left[\frac{\eta (\varepsilon_1)\eta (\varepsilon_2)}{\eta (\varepsilon_0)}\frac{\eta^+_u}{\eta_l}\right]
\end{equation}
The terms in the first bracket are:
\begin{equation}\label{eq:a2.28}
\left[\ldots\right] = \left[N^+_u N (\varepsilon_1) N (\varepsilon_2) v_1 v_2 \sigma^\text{rec} - N_l N (\varepsilon_0) v_0 \sigma^\text{ion}\right]\,\delta(\{\varepsilon\}) d\varepsilon_0 d\varepsilon_1 d\varepsilon_2
\end{equation}
where $\delta(\{\varepsilon\}) = \delta (\varepsilon_0 - \varepsilon_1 - \varepsilon_2 - I_l)$ is necessary for energy conservation. From the detailed balance relation (\ref{eq:Fowler}), the first term in (\ref{eq:a2.28}) can be written as:
\begin{align}
N^+_u N (\varepsilon_1) N (\varepsilon_2) v_1 v_2 \sigma^\text{rec}
&= \sigma^\text{ion} \frac{g_l}{g^+_u} \left(\frac{h^3}{8\pi m_e^2}\right)\,\left(\varepsilon_1\varepsilon_2\right)^\frac{1}{2}\left(\frac{\varepsilon_0}{\varepsilon_1\varepsilon_2}\right) N^+_u N (\varepsilon_1) N (\varepsilon_2) \nonumber \\
&= \sigma^\text{ion}\,\frac{g_l}{g^+_u} \left(\frac{2\varepsilon_0}{m_e}\right)^\frac{1}{2}
\left[\left(\frac{8\pi m_e^2}{h^3}\right)\left(\frac{2\varepsilon_0}{m_e}\right)^\frac{1}{2}\right]\,N^+_u \eta (\varepsilon_1)\eta (\varepsilon_2)
\end{align}
where $\eta_p \equiv \eta (\varepsilon_p)$, $p=0,1,2$. We can now re-write (\ref{eq:a2.28})as:
\begin{equation}\label{eq:a2.30}
\left[\ldots\right] = \sigma^\text{ion}\left[g_l g(\varepsilon_0)v_0\right]\,\left[\frac{N^+_u}{g^+_u}\eta (\varepsilon_1)\eta (\varepsilon_2)-\frac{N_l}{g_l}\eta (\varepsilon_0)\right]\,\delta(\{\varepsilon\}) d\varepsilon_0 d\varepsilon_1 d\varepsilon_2
\end{equation}
where $g(\varepsilon_0) = \mathcal{G}(\varepsilon_0)/ \mathcal{V}$. Thus, the rate of entropy change (\ref{eq:a2.27}) becomes:
\begin{equation}\label{eq:a2.31}
\frac{\partial }{\partial t} \delta S = k \,\sigma^\text{ion}\left[g_l g(\varepsilon_0) v_0\right]\,\left[\eta^+_u \eta (\varepsilon_1)\eta (\varepsilon_2)-\eta_l\eta (\varepsilon_0)\right]\cdot\ln\left[\frac{\eta (\varepsilon_1)\eta (\varepsilon_2)}{\eta (\varepsilon_0)}\frac{\eta^+_u}{\eta_l}\right]\,\delta(\{\varepsilon\}) d\varepsilon_0 d\varepsilon_1 d\varepsilon_2
\end{equation}
This rate is of the form $\frac{\partial }{\partial t} \delta S \propto (x-y)\ln\left(\frac{x}{y}\right)$ with all other factors being positive quantities, this expression is always positive when either $x>y$ or $x<y$, and is zero when $x=y$. Thus, the entropy always increases during relaxation and equilibrium is reached when
\begin{equation}\label{eq:a2.32}
\eta_l\eta(\varepsilon_0) = \eta^+_u \eta(\varepsilon_1)\eta(\varepsilon_2)
\end{equation}
This result is an expression of the H-theorem for the collisional ionization/recombination process. 
Note that in order to verify the H-theorem we need mass {and energy conservation} and detailed balance, itself a result of microscopic reversibility (i.e. time-reversal).

A similar procedure can be used for the excitation/deexcitation process. In that case,the rates of excitation and deexcitation can be written as, for a transition $l\leftrightarrow u$:
\begin{subequations}\label{eq:a2.33}
\begin{align}
dR^\text{exc} &= N_l N(\varepsilon_0)  v_0 \, \sigma^\text{exc} d\varepsilon_0\\
dR^\text{dex} &= N_u N(\varepsilon_1)  v_1 \, \sigma^\text{dex} d\varepsilon_1 
\end{align}
\end{subequations}
Note that $\varepsilon_0-\varepsilon_1 = \Delta E_{lu}$ due to energy conservation. Similarly to the case of ionization/recombination (\ref{eq:a2.26}), the rate of change in entropy is:
\begin{equation}\label{eq:a2.36}
\frac{\partial }{\partial t} \delta S = k \,\left[dR^\text{dex}-dR^\text{exc}\right]\cdot\ln\left[\frac{\eta(\varepsilon_1)}{\eta(\varepsilon_0)}\frac{\eta_u}{\eta_l}\right]
\end{equation}
Using the Klein-Rosseland relation, this becomes:
\[
\frac{\partial }{\partial t} \delta S \propto  \sigma^\text{exc} \cdot \left[
\frac{N_u }{g_u} N(\varepsilon_1)\sqrt{\varepsilon_1}\,g_l\left(\frac{\varepsilon_0}{\varepsilon_1}\right) 
- N_l N(\varepsilon_0)\sqrt{\varepsilon_0}
\right]
\ln\left[\frac{\eta(\varepsilon_1)}{\eta(\varepsilon_0)}\frac{\eta_u}{\eta_l}\right]
\]
or
\begin{equation}\label{eq:a2.37}
\frac{\partial }{\partial t} \delta S \propto g_l \varepsilon_0 \, \sigma^\text{exc}\cdot\left[\eta_u\eta(\varepsilon_1)-\eta_l\eta(\varepsilon_0)\right]
\ln\left[\frac{\eta(\varepsilon_1)}{\eta(\varepsilon_0)}\frac{\eta_u}{\eta_l}\right]
\end{equation}
This result is also of the form $(x-y)\ln\left(\frac{x}{y}\right)$, implying that the entropy increases during relaxation until equilibrium is satisfied by the relation between occupation numbers:
\begin{equation}\label{eq:a2.38}
\eta_u \eta(\varepsilon_1)=\eta_l \eta(\varepsilon_0)
\end{equation}

\end{appendices}

\end{document}